# *Comparative Analysis of Search Approaches to Discover Donor Molecules for Organic Solar Cells*

Mohammed Azzouzi[1,2,*], Steven Bennett[1], Victor Posligua[1], Roberto Bondesan[3], Martijn A. Zwijnenburg[4], Kim E. Jelfs[1,*]

1. *Department of Chemistry, Imperial College London, White City Campus, W12 0BZ, London* United Kingdom.

2. *Laboratory for Computational Molecular Design (LCMD), Institute of Chemical Sciences and Engineering, Ecole Polytechnique Federal de Lausanne (EPFL), 1015 Lausanne, Switzerland*

3. *Department of Computing, Imperial College London, London SW7 2AZ, United Kingdom.*

4. Department of Chemistry, University College London, 20 Gordon Street, London WC1H 0AJ, United Kingdom.

* Corresponding authors

Mohammed.azzouzi@epfl.ch, K.jelfs@imperial.ac.uk

## Abstract

Identifying organic molecules with desirable properties from the extensive chemical space can be challenging, particularly when property evaluation methods are time-consuming and resource intensive. In this study, we illustrate this challenge by exploring the chemical space of large oligomers, constructed from monomeric building blocks, for potential use in organic photovoltaics (OPV). For this purpose, we developed a python package to search the chemical space using a building block approach: *stk-search*. We use *stk-search (GitHub link)* to compare a variety of search algorithms, including those based upon Bayesian optimization and evolutionary approaches. Initially, we evaluated and compared the performance of different search algorithms within a precomputed search space. We then extended our investigation to the vast chemical space of molecules formed of 6 building blocks (6-mers), comprising over $10^{14}$ molecules. Notably, while some algorithms show only marginal improvements over a random search approach in a relatively small, precomputed, search space, their performance in the larger chemical space is orders of magnitude better. Specifically, Bayesian optimization identified a thousand times more promising molecules with the desired properties compared to random search, using the same computational resources.

# Introduction

Organic semiconductors have emerged as a versatile class of materials, holding promise for various optoelectronic applications, including in flexible screens, electronic devices, and transparent, lightweight photovoltaic systems.[1,2] However, the successful adoption and integration of organic molecules into targeted devices heavily relies on the discovery of new molecules with optimal optical and electronic properties, as well as considerations of their synthesis cost, solubility in green solvents, and chemical and physical stability.[3]

Exploring the vast chemical space of molecules for organic electronics presents a significant challenge. With an abundance of different molecular structures available for investigation, even slight changes in the chemical composition can profoundly impact the properties of these materials. Among the various approaches to explore this chemical space, a building block strategy is highly attractive.[4,5] By constructing larger molecules from smaller building blocks, we gain the ability to define a chemical space solely based upon combinations of these building blocks.[6] This combinatorial definition of the chemical space renders it more manageable for exploration. Thus, the chemical space can be enumerated and is constrained by the size of the building block library and the number of building blocks in the oligomer molecule. With the defined chemical space, the next step is to evaluate the potential of the molecules for the targeted application. Ideally, we would determine a molecule's properties by synthesising the molecule in the laboratory and measuring its characteristics. This step is time and resource expensive, and unfeasible at a large scale considering the size of the chemical space. To reduce the cost of the search, we can use computational evaluation to determine a smaller number of potentially promising molecules.

A computational evaluation requires two steps: assembling the building blocks to construct a molecular model, and a second step in which the properties of the molecule are predicted using computational chemistry methods. Several tools are available to build molecules from building blocks, offering good starting geometries for the constructed molecules.[7] Specifically, we consider in this work for this purpose our software package *stk*, which offers automated assembly and geometry optimisation.[8] The next step is to evaluate the potential of the molecule for the target application. In the literature we can distinguish between property based evaluation functions, which directly relate to relevant properties of the molecule such as optoelectronic properties (e.g. excited state energy, ionisation potential),[9] and accessibility based evaluation functions,[10] that focus on the synthesisability of the molecule and its ease of use for the application of interest. For example, in the case of organic electronic, we are interested in how easily we can deposit the molecule on a surface to form a film.

Evaluating optoelectronic properties typically requires computationally expensive quantum chemistry calculations that can take hours to days.[11] Consequently, a brute force search of the entirety of the possible chemical space quickly becomes unfeasible. We therefore require efficient search strategies for navigating the vast chemical space to find the most promising systems. One approach that has been explored is the development of machine learning models that alleviate the use of expensive quantum chemical calculations.[12-16] These models can be used as an initial filter in a high-throughput approach to reduce the size of the chemical space of interest to a more manageable size.[14,17] The application of statistical models for molecular discovery is, however, limited by the availability of representative datasets upon which to build the statistical model. This limitation can result in statistical models with low accuracy and biased predictions, which could hinder the discovery effort. Another approach relies on the use of adaptive strategies, which selectively explore the search space, and suggest the most promising candidates based on prior knowledge.[18] These adaptive strategies often incorporate domain-specific information, historical data, or heuristics to guide the search process effectively. Evolutionary algorithms, as an example, demonstrate the power of adaptation in optimization. These algorithms mimic the process of natural selection, iteratively improving candidate solutions to complex problems. By combining variation, selection, and adaptation, they explore the search space effectively.[19, 20] For instance, Greenstein *et al.* employed an evolutionary algorithm, leveraging specified building blocks, to computationally explore the space of potential organic molecular acceptors and donors specifically for organic solar cell applications.[5]

Bayesian optimization (BO) is another powerful approach for optimizing complex, expensive-to-evaluate functions. Unlike evolutionary algorithms, which explore the search space through variation and selection, BO leverages probabilistic models to guide the search efficiently. Specifically, it employs a cheap to evaluate surrogate model that approximates the target property of the search strategy and encodes uncertainty about it. Leveraging this information, the system identifies the next optimal candidate for evaluation based on user-defined criteria. BO has gained prominence as an effective approach for guiding chemical and material discovery. BO's advantages lie in sample efficiency, flexibility, and versatility.[21] For example, Strieth-Kalthoff *et al.* recently used BO to explore an enumerated space of organic molecules for laser applications, showing a considerable improvement in the search efficiency compared to other approaches.[22-24] When implementing BO for chemical or molecular discovery, the user faces considerable challenges related to the choice of different molecular representation options and the high dimensionality of the representation space. Molecular representations vary widely, from traditional descriptor-based vectors and molecular fingerprints (e.g., Mordred, ECFP) to string-based formats like SMILES, graph-based embeddings used by GNNs, and even grid or image-based 3D encodings, and each representation comes with distinct trade-offs in terms of interpretability, invariance properties, and computational cost.[25] Moreover, in BO we define a decision criterion in the form of an acquisition function to determine which point in the search space should be evaluated next. The acquisition function balances the exploration-exploitation trade-off: exploring regions of uncertainty (where the surrogate model is uncertain about the fitness), while also exploiting promising areas (where the surrogate model predicts high fitness).[26] The optimisation of the acquisition function over the discrete spaces that are particularly relevant in chemical discovery is very challenging.[21, 27]

      Here, we introduce a Python package, *stk-search*, that can execute a variety of search algorithms within a molecular chemical space. We explored the application of this package and different search algorithms for a use case targeting organic molecules for application in OPVs. We first evaluated and compared the performance of the different search algorithms on a benchmark dataset of precomputed search space (comprising 30,000 different oligomers) using a variety of metrics. Then, we investigated how the performance extends to searching across the vast chemical space of 6-mers (comprising over $4 \times 10^{14}$ oligomers built from 6 constituent building blocks). Finally, we analysed the new oligomers and compared them to oligomers present in the benchmark dataset.

# Methods

We first summarise here the overarching search strategy employed in *stk-search*, followed by a description of the distinct search algorithms implemented in the package.

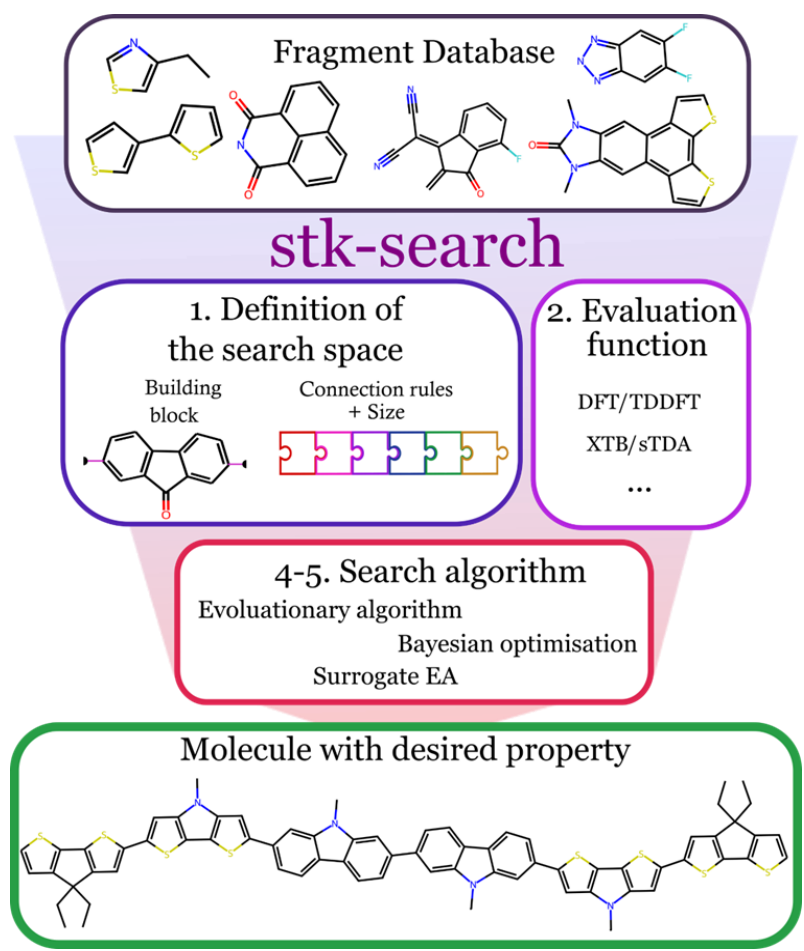

*Figure 1 Overview of stk-search. Summary of the procedures implemented in stk-search to explore the chemical space of molecules constructed from building blocks. Starting from a fragment database, we first define the chemical space by generating building blocks with specific connection points, and then establish the size and the connection rules to build the molecules (1). Next, we define an evaluation function where we build the molecules and evaluate their properties using quantum chemistry methods (2). Finally, using an appropriate search algorithm, we can explore the chemical space (4-5) and find molecules with the desired property.*

1. ***stk-search* overview**

We developed *stk-search*, an open-source Python package, to efficiently search the chemical space of molecules constructed from smaller building blocks. The package leverages our existing *stk* software, used to assemble the models of the molecules,[8] along with the *BoTorch* package[28] for Bayesian optimization and *PyTorch*[29] with *Torch Geometric* for neural network models.[30] *stk-search* offers Python functions to facilitate the calculation of the molecules' properties using quantum chemistry calculations. The resulting molecular geometries or position matrices are stored in a MongoDB database, alongside the results of the property predictions.

The approach used to search the chemical space within *stk-search* can be summarised by the following steps (Figure 1):

1. Definition of the chemical search space of the constructed molecules to be explored through the choice of building blocks, the number of building blocks, as well as connection rules for the formation of the larger molecules. Here, the building blocks are molecular fragments with predefined connection points and connection rules. (supporting information 1.a for more details on the search space definition)
2. Establishing an evaluation function that the search algorithm will seek to maximize or minimize in a target molecule. This function can be either a single property, or a combination of properties.
3. Selecting an initial population of candidate molecules from the defined chemical space, using user-defined criteria or random or pseudo-random sampling.
4. Constructing the molecules and evaluating their properties before adding predicted structural and property information for these molecules to the stored database.
5. Using a search algorithm to suggest new molecule(s) to evaluate.
6. Repeating steps 4 and 5 for a user-defined number of iterations or until the computational budget has been exhausted.

## 2. Background of the implemented search algorithms

For the four specific types of search algorithms implemented in *stk-search*, we distinguish first between model-free methods and methods that rely on the use of a surrogate model.

In the case of the model-free methods, we considered two examples:

a) **Random grid search (Rand)**; a simple approach where the molecules evaluated are randomly selected without replacement from the defined searched space. Without replacement here means that once a molecule has been selected it cannot be selected again.

b) **Evolutionary algorithm (EA)**; a derivative-free optimization approach, which explores the vast chemical space following rules mimicking the principles of evolution. One iteration of the EA algorithm consists of, as shown in Figure 2, the following steps: (i) we select an initial population of pre-evaluated parent-molecules based on their evaluation function (often referred to as a 'fitness function' in the context of EAs); (ii) from this parent population, a new population of offspring molecules is generated using mutation and crossover operations involving the building blocks; (iii) one or several candidates are randomly selected for evaluation from within the population of offspring molecules. These steps are then repeated for a set number of iterations or until a predefined convergence criteria is reached. While EAs can be powerful approaches to significantly reduce the number of molecules that need to be evaluated before identifying optimal molecules, optimisation of the EA's parameters to increase its efficiency involves adjusting many parameters, including the number of crossovers, mutations, parents, and the number of molecules suggested for evaluation after each iteration.[5, 20, 31, 32] This parameter optimisation is particularly challenging when the evaluation function is expensive to evaluate, as in the cases relevant to chemical discovery. For this reason, model-based methods are more attractive.

For the model-based methods, we considered two different approaches: methods that use the prediction of a surrogate model without information related to the uncertainty (this is considered a greedy approach), and methods that incorporate a measure of uncertainty in their approach.[33] Here, uncertainty refers to the variability and potential error associated with the prediction made by the surrogate model. The two model-based methods are:

c) **Surrogate Evolution algorithm (SUEA)**; a greedy approach that uses efficient surrogate models to approximate the evaluation function and improve the performance of the EA.[34] We use a surrogate model trained on previously evaluated molecules (before the start of the search campaign) to select a molecule in the offspring population to be evaluated. At each iteration of the search algorithm (Figure 2), a new parent population is chosen, and the pretrained surrogate model is used to identify the most promising molecule in the offspring population. The molecule with the best predicted property (evaluated using the pretrained surrogate model), will then be evaluated using the chosen evaluation function (expensive evaluation function, e.g., quantum chemistry calculation) and will be added to the population of potential parents.

d) **Bayesian optimisation (BO)**; The second category of model-based methods are methods that use the uncertainty of the value predicted by the surrogate model to guide the selection of molecules to evaluate. In this context, the surrogate model provides an estimate of the evaluation function and predicts the uncertainty associated with that estimate. BO transforms the optimization problem from a costly-to-evaluate black-box function to an acquisition function that is easier to optimize. An example of an acquisition function is the sum of the predicted value and its uncertainty (usually a multiple of the standard deviation or variance of the predicted value), known as the upper confidence bound (UCB). When the UCB serves as the acquisition function, selecting a potential molecule depends not only on the predicted evaluation function but also on our confidence (or uncertainty) in that prediction. One iteration of BO within *stk-search* consists of the following steps: i) Train a surrogate model using Gaussian processes on all or a subset of the evaluated molecules in the search space; ii) Find the molecule(s) in the search space with the highest acquisition function; iii) Evaluate the molecule(s) and add them to the list of molecules the Gaussian process will be trained on. The search algorithm is run for a set number of iterations or until the computation budget is exhausted. For the acquisition function, we can use several acquisition functions implemented in BoTorch, such as expected improvement and UCB. The expected improvement (EI) acquisition function measures how much better a potential solution is expected to perform compared to the current best solution. The optimisation of the acquisition function over the space of molecules is a challenging endeavour. As the acquisition function is a quick to evaluate function, we use an EA to optimise it. In each iteration of the BO search algorithm, we optimize the acquisition function using an EA. The EA runs for multiple iterations until it converges, and we consider many (in the order of thousands) molecules per generation. The use of the EA here avoids the need to evaluate the acquisition function across the entire chemical space, which is infeasible due to the vast number of molecules.

3. **Surrogate models**

For the two model-based methods, we consider models that relate a mathematical representation of molecules to their desired property. The model is trained on prior evaluation of the molecules in the search space. For SUEA, since the model is pretrained before the search process, we can utilize any available model if the inference cost is manageable. This means we can employ traditional machine learning models like random forests for specific molecular representations, or graph neural networks that leverage the position and nature of atoms to construct a representation.[35, 36]

For BO, it is essential to use a surrogate model that can be trained quickly and provides an uncertainty measure. Therefore, Gaussian processes are commonly preferred for BO.[21] The use of Gaussian processes for molecular systems requires the representation of molecules as mathematical objects, such as arrays or graphs.[37-39] The choice of such representation can strongly influence the performance of the search algorithm, and this choice can be done prior to the search by analysing the existing data we have for the search space. We distinguish here between constructed molecular representations built from a chosen set of properties or molecular descriptors of a molecule's building blocks and learned molecular representations from data available prior to the search campaign. Apart from the choice of representation, for the Gaussian process, the user needs to choose among different kernels that define how the similarity between two molecules relates to the target property. The similarity between two molecules is a function of the representation used, which in this case is often an array representation of the molecules. The different kernels currently implemented in *stk-search* are Mattern, Tanimoto and radial basis function.[39, 40]

In *stk-search*, we have incorporated the ability to train and utilize surrogate models based on graph neural networks (GNNs). GNNs are powerful models for learning representations from graphical data, making them well-suited for modelling molecular systems. GNNs operate by iteratively updating node features using message-passing operations within the graph structure. The models considered here are 3D based models such as SchNet that take the position matrix of the atoms forming the molecules as input and predict a scalar property of the molecule.[12, 41 15, 41, 42] Our implementation relies on the implementation of a graph neural network by Liu *et al.* in their package *Geom3D*.[15] Since molecules are defined by building blocks and assembly constraints, their atom position matrix is not immediately accessible. To address this, we employ *stk* to assemble the molecules and create an initial geometry. We then use this generated position matrix as input for our model. The initial geometry step is efficient and parallelizable, ensuring it does not impact the search algorithm's performance. Additionally, when molecular geometry significantly influences the evaluation function, we incorporate a training step that relates the specific quantum calculation's geometry to the one initially generated by *stk*.

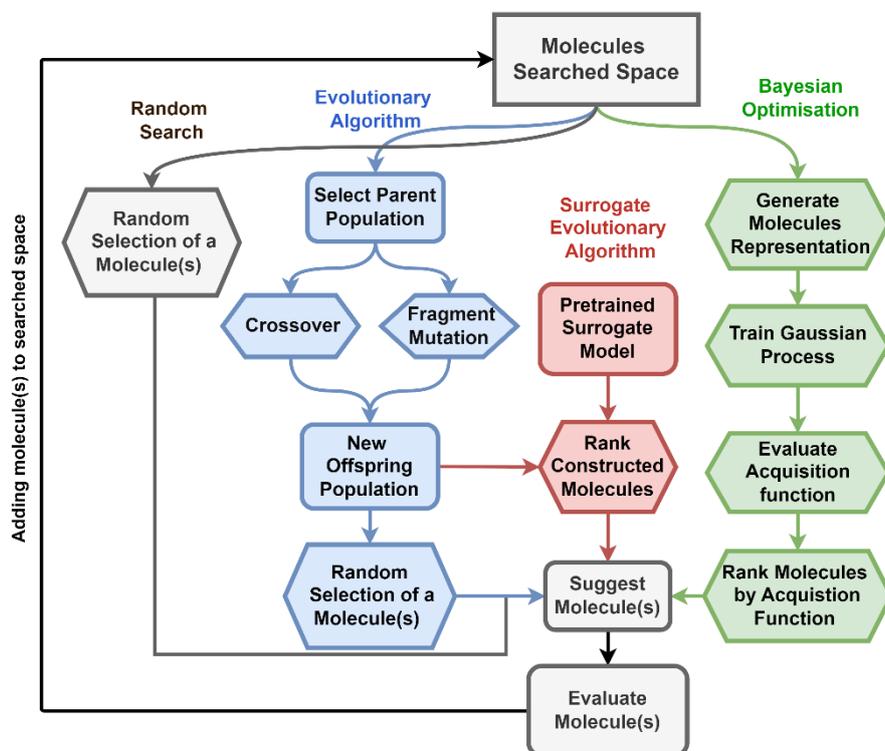

*Figure 2. Diagram representation of the different search algorithms implemented in stk-search. The different search algorithms include an evolutionary algorithm (EA), surrogate EA (SUEA), and Bayesian optimisation (BO).*

# Results and discussion

## 1. Search space definition

We used *stk-search* on a specific use case; exploring the chemical space of oligomers formed of 6 building blocks, representative of the oligomers and polymers in organic semiconductor applications.[5] For example, the non-fullerene acceptors used in OPV applications can be split into 3 to 5 different constituent building blocks: ADA or ADA'DA, where A is an electron deficient unit and D an electron rich unit.[43] For the donor molecules, they can be complex copolymers for which the unit cells can be split into 4 to 6 building blocks.[44] The chemical space considered here would cover both the space of donors and acceptors currently used and much more. Without introducing any conditions on the building blocks and their positions in the molecule, the number of molecules in the chemical space is $N^6$, where $N$ is the number of unique building blocks considered.

As a test case with relevance to the broader organic electronics field, we specifically sought donor oligomers that would work efficiently in a single layer bulk-heterojunction device with the most efficient acceptor in the field (namely Y6[44]). We chose here to focus on donor oligomers formed of 6 building blocks as a compromise between loss of accuracy in shortening the oligomer, and the increased cost of screening larger systems. Prior work by some of us showed that the optoelectronic properties of interest here converge with oligomer chain lengths of 6 monomers.[45] The compatibility of the polymer with Y6 requires the donor oligomer to have an ionisation potential (IP) 0.1 to 0.2 eV higher than Y6 (which is around 5.65 eV as experimentally measured[46]) to reduce the energy losses related to the exciton dissociation process and ensure high charge generation yield.[47] The donor should also absorb strongly in the spectral region where Y6 absorbs little or no light (in the spectral region from 400 - 550 nm). These oligomer properties can be calculated using density functional theory (DFT), and time-dependent density functional theory (TD-DFT).[48, 49] Specifically, we can calculate the vertical ionisation potential of a single oligomer in vacuum as the difference in ground state energy between the neutral oligomer and its positively charged version. For the optical properties, we limit our calculation to the properties of the first vertical excited state using TD-DFT calculations. We calculate the energy of the first excited state ($E_{s1}$) as a proxy for the spectral region where the molecule would absorb, and the oscillator strength of the transition from ground state to first excited state ($f_{osc,S1}$) as a proxy for the strength of the transition (*i.e.* the absorption coefficient).[50]

We used the *stk*-generated geometries as initial input, then used the Experimental-Torsion basic Knowledge Distance Geometry (ETKDG) approach in *stk/RDKit* to generate a first geometry.[51] Next, we optimised the geometry of the lowest energy conformer found using GFN2-XTB[52] and calculated the vertical ionisation potential and electron affinity using the IPEA option in XTB. The optical properties of the oligomers were calculated using sTDA-XTB.[53] The calculated properties using this level of theory can be related to the experimentally measured properties using a linear transformation.[54] This level of theory was chosen because it provides a good balance between computational efficiency and accuracy, making it suitable for the high-throughput screening of potential donor molecules for OPV applications.[45] A higher level of theory or , indeed, experiments can be used to evaluate the most promising candidates further, but this is out of the scope of the paper, as our main focus is on the comparison between the different search approaches.

To create an evaluation of potential oligomeric molecules that considers the factors mentioned above, we used the following evaluation function that is a combination of the three properties (IP, $E_{s1}$, $f_{osc,S1}$):

$$F_{comb} = -|E_{S1} - 3| - |IP - 5.5| + log_{10}(f_{osc,s1}) \qquad (1)$$

We will refer to the value of the evaluation function (equation 1) for a molecule as the combined property function ($F_{comb}$) of the molecule. The ideal IP is set to 5.5 eV, and the target excited state energy to 3 eV (~410 nm). The oscillator strength in this case should be maximised. A value of $F_{comb}$ above zero means that we have molecules with IP and $E_{s1}$ close to the target, and an $f_{osc,s1}$ above 1. An oscillator strength above 1 can be related to an absorption coefficient of the film of a value ~0.02 nm$^{-1}$ (depending on the arrangement of the molecules and other parameters), meaning a film of a thickness of ~50 nm would absorb all the light at that wavelength.[55] In the case where $f_{osc,s1}$ is zero, indicating a dark first excited state which is detrimental for the use of the molecule as donor in an organic solar cell; the overall score of the molecule in this case is set to a low value of -10. We provide in the supporting information 1.e details on the computational implementation of the evaluation function.

The chemical space considered in this example, consists of 131 different fragments from the library of Greenstein *et al.*[5], these are shown in Figure S1. We limited the number of atoms per fragment to 30 non-hydrogen atoms. The library of fragments can be combined into building blocks and becomes a library of N=274 different building blocks when all possible routes to combining the fragments are considered. We manually clustered these building blocks by chemical similarity and representative molecules for each cluster are shown in Figure 3a (Figure S9 shows the different clusters in 2-dimensional space). These clusters would help us analyse the overall performance of the molecules suggested by the different search algorithms. Cluster 0, for example, is formed of building blocks similar to 3-(dicyanomethylidene)indan-1-one (2HIC), which is an electron withdrawing end-group commonly used to prepare non-fullerene acceptors.[56] Whereas cluster 4 is formed of three fused-ring building blocks such as fluorene derivatives, which are commonly used in polymer semiconductors.

All ways of combining the 274 building blocks presented above creates a chemical space of $N^6 > 10^{14}$ different 6-mers. In what follows, we first assess the performance of 6 different search algorithms on a constrained chemical space limited to 30,000 randomly precalculated 6-mers. Then we used the different search algorithms to search the larger $10^{14}$ chemical space.

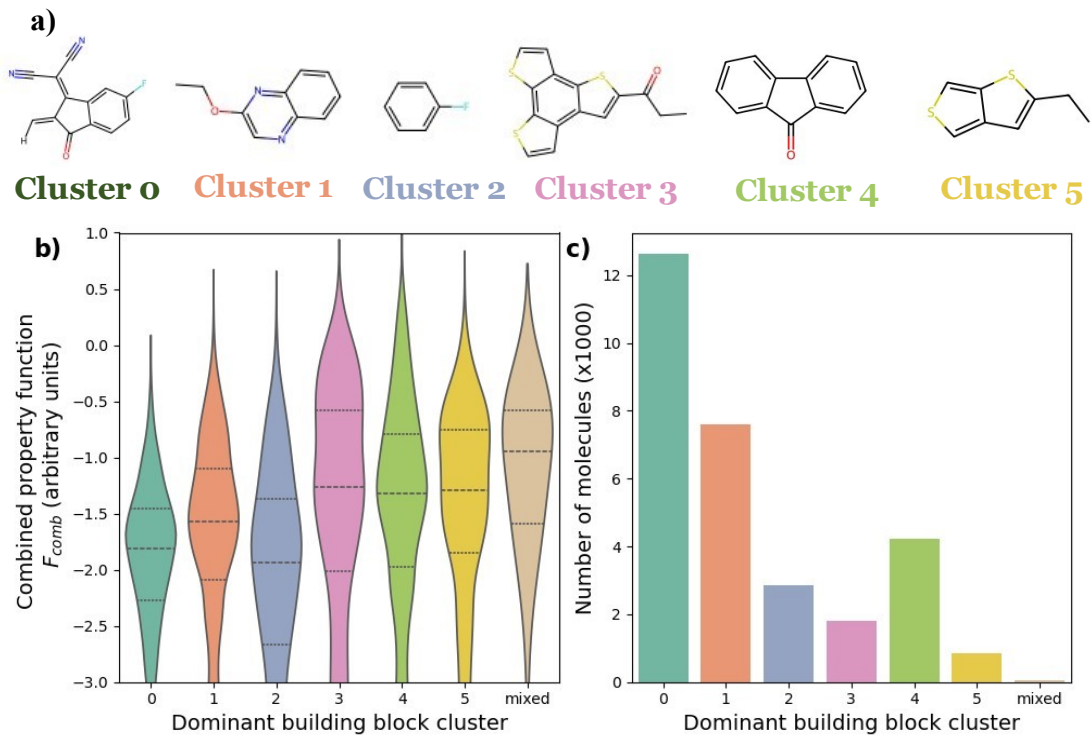

Figure 3. Features of the precalculated search space of 30,000 6-mer molecules; a) Different representative building blocks for 6 different clusters of oligomers; b) Violin representation of the distribution of $F_{comb}$ values of the oligomers with dominant building blocks from different clusters. Here we defined the "dominant building block cluster" as the most frequent building block in the molecule, and when there is no dominant building block cluster, we classify the constructed molecules as "mixed"; the dashed lines split the distribution into 4 quartiles. c) the frequency of molecules in each cluster of dominant building block clusters.

## 2. Implemented search approaches

Using the four distinct search algorithms described in Section 2, we applied them here with different implementations to create six distinct search approaches. These include the four search algorithms *Rand*, *EA*, BO and *SUEA*. For the BO we consider three different implementations that differ in the representation and surrogate model considered. Here, we consider three different representations to investigate the impact of choosing a molecular representation on the search algorithm performance. Specifically, the six search approaches considered are:

1. Random search (*Rand*): used as our baseline case.
2. Evolutionary algorithm (*EA*); we applied a simple case where five parents are chosen for each generation. Two of the parents are chosen randomly, and the other three are taken as the molecules with the highest $F_{comb}$ in the current population. Next, we consider all the mutation and crossover operations possible to generate an offspring population and randomly select a molecule in the offspring population to evaluate. See supporting information 1.d for more details.
3. Surrogate EA (*SUEA*); We used the same approach as the EA in (2) but using a pretrained model to select the best molecule in the offspring population, rather than randomly choosing one. The pretrained model used here is a deep neural network that relies on the architecture of SchNet.[12] SchNet was selected due to its well-demonstrated balance between computational efficiency and representational power, making it a practical yet effective choice for large-scale screening tasks. We use it to generate an array representation of the molecules that takes as input the atom types and their coordinates in a specific geometry. This representation is then passed through a feed forward neural network to predict $F_{comb}$. Since the properties considered here are strongly affected by the geometry of the molecules considered, we aim to generate a representation of the molecule that is related to their optimised geometry in the ground state at 0 K, i.e. the geometry generated following the optimisation using xTB. However, we want to avoid running relatively expensive geometry optimisations to predict the property of interest using the surrogate model. Hence, we include a simple neural network in the training process to map the representation generated using SchNet with the position matrix of the molecules from the *stk*-generated geometry to the representation generated with the xTB optimised geometry. The model is trained on a subset of the precalculated molecules in the database. See supporting information 1.d more details about the SUEA implementation, and supporting information 1.e for details on the implementation of the surrogate model used.

4. Bayesian optimisation with a representation generated using the optoelectronic properties of the molecules building blocks (*BO-Prop*). We considered this representation to have potential benefit given that the optoelectronic properties of the larger molecules are related to optoelectronic properties of the building blocks.[25] Here, we specifically consider a list of the properties of the building block calculated using xTB and sTDA-xTB; IP, $f_{osc,s1}$ and $E_{S1}$.[52, 53]

5. Bayesian optimisation with Mordred descriptors of the building blocks (*BO-Mord*). This considers more general descriptors that are not limited to optoelectronic properties of the building blocks. We built a lower dimensionality representation of the 1200 descriptors available in the Mordred program for each building block and concatenated them to form an array representation of the molecule.[57]

6. Bayesian optimisation with a learned representation (*BO-Learned*), where we used a data driven approach to learn a relevant representation for the property of interest using prior generated data (*i.e.* data generated from previous exploration of the search space and stored in the database). We used the same neural network as for the surrogate model presented in the SUEA in (3) to generate a representation of the molecule. This approach is similar to using a deep kernel to describe the similarity between the molecules for the Gaussian process.[58] Using this learned representation, we aimed to investigate how the search algorithm would be affected if we used a representation inferred from fitting prior data. Further details are in supporting information 1.d and 1.e. This approach addresses the limitation of Gaussian processes when dealing with large datasets. It achieves this by using a molecular representation that has been learned from a larger number of training molecules. This representation improves the performance of the Gaussian processes without them needing to be trained on the same number of molecules.[59]

The selection of parameters for various search algorithms can introduce considerable bias, affecting the performance and outcomes in molecular discovery. For example, in an *EA*, the choice of parents and the types of mutation and crossover operations can direct the search towards specific regions of the chemical space, potentially neglecting other promising areas. Similarly, in BO, the selection of surrogate models and molecular representations can result in biased predictions. Furthermore, the use of pretrained models in *SUEA* and *BO-Learned* involves another set of hyperparameters that can significantly influence the search results. Given the complexity of evaluating the overall performance of a search algorithm for a particular task, as discussed in more detail in the following section, we limited our choice of search algorithm parameters to a specific set established through a non-exhaustive parameter exploration.

## 3. Assessing the performance of the search algorithms

Assessing the performance of a search algorithm and approach on unknown chemical space is challenging due to the considerable number of parameters to consider. Different search algorithms can perform better or worse for different tasks, and it is often hard to predict their performance on unknown space *a priori*.[60-62] Here, the aim of our search campaign was to find new molecules with target properties above a threshold with the least number of quantum chemical calculations having to be performed, given that the quantum chemical calculations are the bottleneck for the high-throughput exploration, and have the largest resource cost.

We first describe the performance of the search algorithm on a restricted benchmark space where we limit the chemical space to 30,000 molecules that we had previously calculated. Running the search on a benchmark where we know the best solutions helps us to assess how well the search approaches perform. Then we assess the performance of the search algorithm when run over the space with more than $10^{14}$ molecules.

### 3.a. Benchmark comparison of search algorithm performance

The initial precalculated benchmark space, comprising 30,000 molecules, was selected randomly from the total search space of >$10^{14}$ molecules. Figure S12 shows a 2D projection of the chemical space using principal component analysis (PCA) and demonstrates that the precalculated chemical space is diverse and samples across the total search space. Figure 3b shows that no particular building block cluster dominates for either high or low $F_{comb}$ values. This is expected as the link between the oligomers structure and these properties is more complex than that.[45, 49]

We ran each of the six different search approaches described above (*Rand*, *EA*, *SUEA*, *BO-Prop*, *BO-Mord* and *BO-Learned*) for 400 iterations with an initial random population of 50 molecules. We limited the search to a specific number of iterations to mimic the case where we are constrained by computational resources and can only evaluate a limited number of molecules using the expensive evaluation function.[61] We are interested in evaluating how fast the search algorithms find the top 1% of the molecules in the dataset (300 molecules in our case). For the search algorithms that required pretraining (of the representation for *BO-Learned* and of the surrogate model for *SUEA*), we hid the top 1% molecules from the training and validation datasets, and then pretrained on a random selection of 20,000 of the remaining 27,700 molecules in the dataset (performance of the surrogate model is presented in Section S5). A comparison between the performance of the search algorithm with a smaller training set of 10,000 molecules is shown in the supporting information Section 5. To take into consideration the stochastic nature of the search algorithms, we averaged our results over 25 separate runs starting with different initial populations.

The results are shown in Figure 4. For the six different search approaches, we analysed the best (highest) value of $F_{comb}$ found for any oligomer evaluated up to the current iteration (Figure 4a) and the mean $F_{comb}$ for the oligomers at each iteration up to the current one (Figure 4b). The first metric (shown in Figure 4a) shows how fast the algorithm finds the molecules with the best properties. The second metric (Figure 4b) assesses the overall performance of the search algorithm in suggesting molecules that are better than the average molecule in the search space when compared to the baseline. Compared to the baseline *Rand*, the other five search methods consistently identified molecules with a higher $F_{comb}$ value after only 100 iterations, outperforming the best result obtained by *Rand* after 400 iterations. *SUEA* manages to consistently find molecules among the top 30 molecules (top 0.1% in the dataset) after less than 100 iterations. *BO-Learned* is the second best and shows a similar rise in maximum $F_{comb}$ to *SUEA* in the first iterations, however it only reaches the same maximum value as *SUEA* after 300 iterations. The use of the pretrained representation/model speeds up the performance of the search approaches in finding the best molecules in the dataset. The pretraining also helps the approaches choose molecules with a higher $F_{comb}$ in individual iterations. Examination of the molecules selected in the different runs shows that very similar molecules are being selected across different runs of both *BO-Learned* and *SUEA* for the first 50 to 100 iterations (Figure S18).[63] After the first 100 iterations, *BO-Learned* started suggesting more diverse molecules. For the other search algorithms, *EA* performs better than *BO-Mord* and *BO-Prop* in the first iterations, but then gets stuck in a region of lower performing molecules and fails to consistently find the top 30 molecules after 400 iterations. *BO-Mord* and *BO-Prop* show a slow but consistent $F_{comb}$ improvement over the full 400 iterations, as the surrogate model better learns the search space and begins to perform similarly to *SUEA* and *BO-Learned* after 350 iterations.

Figures 4c and 4d focus on exploring how well the search approaches performed at finding the top 1% (300) of molecules, with Figure 4c showing the number of top 1% molecules found up to a given iteration and Figure 4d showing the discovery rate of the top 1% of molecules, which we calculate as *(number of top molecules found)/(number of iterations)*. All other search approaches outperform *Rand* by these metrics, as expected. *BO-Learned* performed the best in finding the highest number of top molecules after 400 iterations, approximately 35 top molecules found on average. The discovery rate of top molecules was particularly high in the early iterations for both *SUEA* and *BO-Learned*, before falling over the course of the searches, suggesting the learned representation/model were performing well from the outset. For *BO-Prop* and *BO-Mord*, the discovery rate increases slowly in the first 100 iterations, then it drops slightly later. By the end of 400 iterations, *BO-Mord* and *BO-Prop* find as many top molecules as *SUEA*.

Ideally, you would be able to complete a search such that the top solutions were found regardless of the initial population. This is not yet the case for the 400 iterations here, and some search approaches, in particular *EA*, show much greater variance of outcome dependent on the 25 different initial populations (as exemplified by wider shaded areas in Figure 4). This emphasises how more effective methods to seed the initial population could significantly improve the search performance.

We extended the evaluation of the search strategies to 800 iterations, allowing each algorithm to explore a larger portion of the chemical space: exceeding 2% of the total benchmark. Compared to the 400-iteration results, a key difference emerged: the random search (Rand) consistently outperformed the model-based approaches after around 700 iterations, identifying molecules with higher $F_{comb}$ scores (see supporting information section 4.b). This outcome underscores the growing impact of dataset-induced biases over longer search horizons. In particular, model-based algorithms such as SUEA, and BO are increasingly constrained by the structural biases in the dataset, favouring molecules composed of frequently occurring building blocks. These biases limit the algorithms' ability to explore under-represented regions of chemical space, especially when the surrogate models and acquisition functions (e.g., Expected Improvement) prioritize candidates that are structurally similar to the majority of the dataset.

. Next, we examined the transferability of these observations to the much more demanding task of searching the unrestricted space of more than $10^{14}$ different 6-mer molecules.

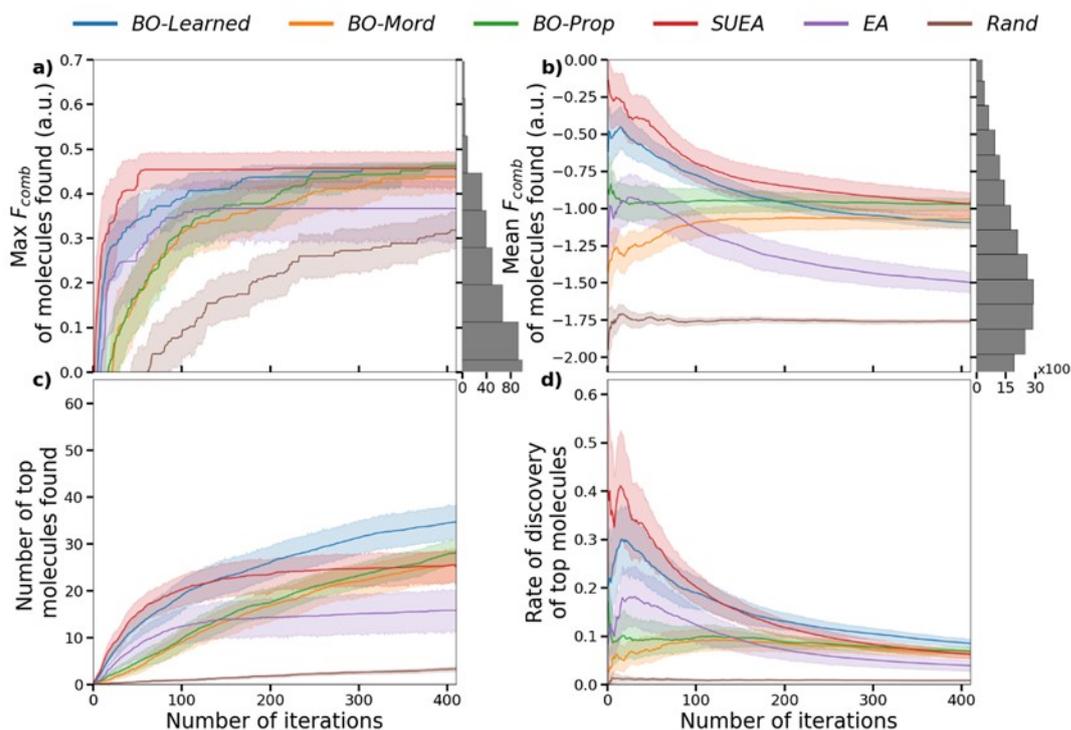

*Figure 4 Performance of the six different search approaches on the precalculated benchmark dataset of 30,000 molecules. The solid-coloured lines show the mean $F_{comb}$ over 25 runs with different initial populations and the coloured shaded area shows the variance of the $F_{comb}$ over those different runs; a) Maximum $F_{comb}$ found for an oligomer up to the current iteration. The histogram on the right shows the distribution of the oligomers in the benchmark dataset; b) Mean $F_{comb}$ of the oligomers found up to the current iteration; c) Number of oligomers in the top 1% found up to the current iteration (top 1% is 300 molecules); d) Discovery rate of the top 1% oligomers in the dataset, calculated as the (number of top molecules found)/(number of iterations).*

### 3.b. Performance of the search algorithms over the full search space

To compare the six different search approaches over the full search space, we ran each approach with a time restriction of 8 hours for a single run, and with the same computational constraints (30 CPUs and 50 GB of memory). The number of iterations performed by each search run depended therefore on the computational time for calculation of $F_{comb}$ for molecules, as well as the computational time needed to suggest new molecules to evaluate. We considered 50 independent runs (with different initial populations) using the same six algorithms used in the benchmark. For *SUEA* and *BO-Learned*, the trained model/representation was the same as for the benchmark study. For each search run, we started from an initial random population of 290 molecules, to which we added the best 10 molecules in the precalculated benchmark space. Adding the best molecules found in the searched space helps ensure that the search approaches start with a better initial population.

Figure 5 shows the distribution of $F_{comb}$ for the new molecules suggested by the different search approaches along the distribution of $F_{comb}$ for the oligomers in the database (in grey). We added the distribution of $F_{comb}$ of the molecules suggested by *BO-Learned* in black in the other plots to facilitate the comparison. First, compared to the molecules present in the benchmark (grey distribution), all the search approaches apart from *Rand* suggested molecules with higher $F_{comb}$. For example, the mean value of $F_{comb}$ for the molecules suggested by BO-Learned is around 0.2, where the mean value of $F_{comb}$ for the molecules in the benchmark was close to -1.9. Second, we find that *BO-Learned* suggested molecules with overall higher $F_{comb}$ compared to the other search approaches. *BO-Learned* suggested the highest ratio of molecules with $F_{comb}$ higher than 0 (69% of suggested molecules), next best by this metric was *BO-Prop* (60%), then *SUEA* (54%).

However, if we explore other metrics to compare the performance of the search approaches, it is a different story. Exploring how the approaches performed at finding molecules with $F_{comb}$ higher than the ones in the initial benchmark dataset, we found that *BO-Mord* suggested the highest number of better performing molecules (16 molecules, Table 1), twice as many as *BO-Learned*. Although *BO-Learned* uses a representation of the molecules that is better to predict their combined property, it fails to find molecules better than the ones in the benchmark. *BO-Prop* and *SUEA* each only found one new molecule better than the ones in the initial benchmark, performing as good as the *EA* that is not model-based approach.

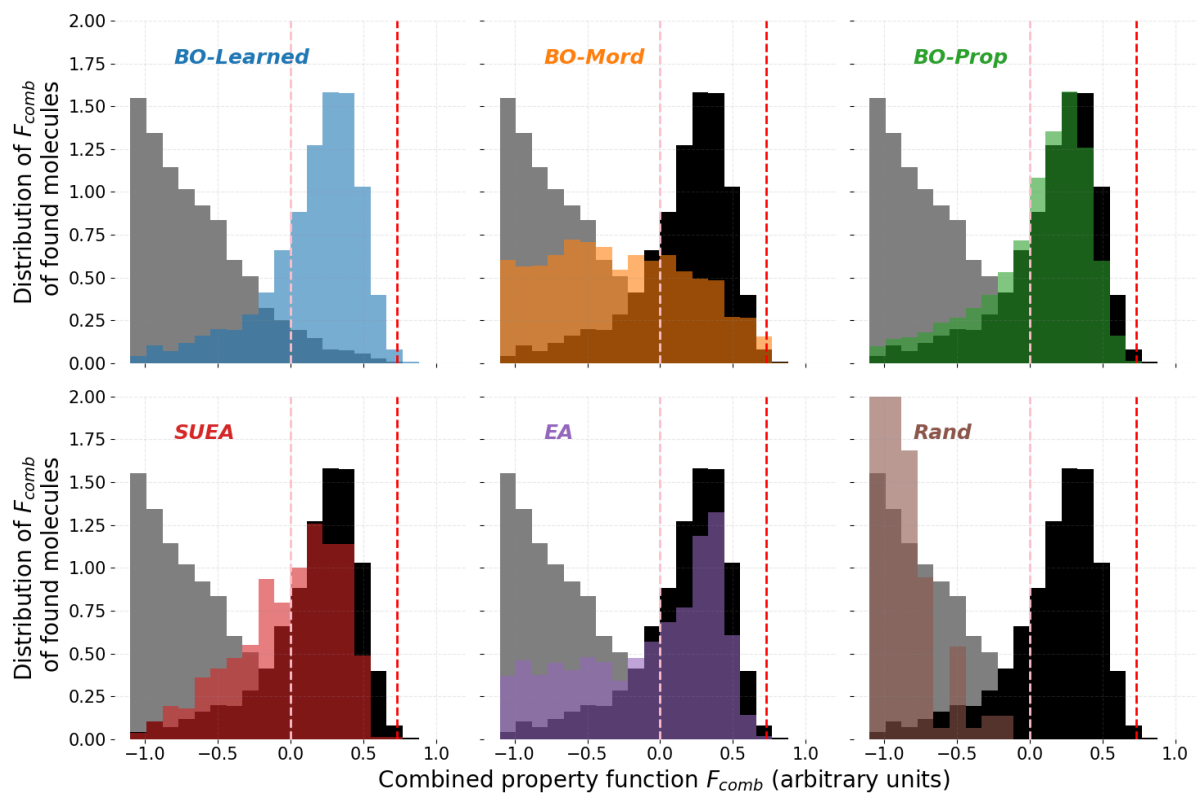

*Figure 5 Distribution of the combined property function ($F_{comb}$) for the molecules suggested by each search approach. The grey distribution shows the distribution of the benchmark dataset. The black distribution shows the distribution from the BO-Learned search algorithm for comparison. The pink dashed line shows the threshold to have target property above 0 and the red dashed line shows the $F_{comb}$ of the best molecule in the benchmark.*

Table 1 Summary of the metrics to compare performance of the different search approaches over the entire search space. Any approaches with pretraining components were pretrained on 20,000 molecules.

| Search algorithm | Unique new evaluations | Unique molecules with $F_{comb} > 0$ | Rate of discovery of molecules with $F_{comb} > 0$ | Molecules found better than benchmark dataset. (outstanding molecules) |
|---|---|---|---|---|
| BO-Learned | 2096 | 1439 | 69% | 8 |
| BO-Mord | 2675 | 535 | 20% | 16 |
| BO-Prop | 3868 | 2336 | 60% | 1 |
| SUEA | 722 | 394 | 54% | 1 |
| EA | 3614 | 1146 | 32% | 1 |
| Rand | 2887 | 0 | 0% | 0 |

To assess the generality of the observations made on the first set of runs, we repeated all the search runs, but this time we trained the models/representations on the total calculated dataset at this point of 58,000 molecules, that is the original 30,000 molecules, along with the 28,000 molecules calculated in the preceding runs. The SchNet model was retrained, and this acts as the surrogate model for *SUEA* and the model to generate the representation for *BO-Learned*. For the other search approach, the new set of runs includes the best molecules found in the 58,000 molecules dataset in the initial population. The $F_{comb}$ distribution of the molecules in the new 58,000 dataset is shown in Figure S26. The performance of the search algorithms in the second set runs is shown in Figure S27 and Table 2. Similar to the first set of results, these findings indicate that the molecules proposed by *BO-Learned*, *SUEA*, and *BO-Prop* have in average higher $F_{comb}$ than the ones suggested by the other search approaches. Additionally, in this set of runs, EA identified the highest number of molecules with $F_{comb}$ higher better than the best molecule in the starting dataset: 7 new molecules. Followed by *BO-Mord* that found 5 new molecules. This further confirms that the improved performance of the surrogate models to predict the value of $F_{comb}$ on the starting dataset reduced the chance of the algorithm to find molecules better than the ones in the starting dataset. This relates to the limitation of the models to extrapolate outside of their training dataset[64].

The results of the search approaches consistently show that the algorithms that use a more accurate surrogate model help the search find molecules with in average higher $F_{comb}$ (*SUEA*, *BO-Learned*, *BO-Prop*). For *BO-Mord*, the algorithm only suggests 20-31% of molecules with $F_{comb}>0$ as compared to 69-54% for *BO-learned*. The representation used for *BO-Mord* did not help distinguish molecules by their $F_{comb}$ value. EA showed a similar performance at suggesting molecules with $F_{comb}>0$ (32 % in the first and second set of runs). However, the two search algorithms *EA* and *BO-Mord*, showed better performance at finding outstanding molecules, i.e. molecules better than the current ones in the starting dataset. This result raises the question of whether a better representation or using a better surrogate model can results in reduced chances of finding molecules better than the ones we had in the starting dataset. This observation is even more important in the case of *SUEA*, as the surrogate model only considers the predicted combined property without any information about its uncertainty. Hence using a pretrained machine learning model can introduce a considerable bias that limits the performance of the search algorithm.

Table 2 Summary of the metrics to compare performance of the different search approaches over the entire search space. Any approaches with pretraining components were pretrained on 58,000 molecules. The starting dataset here refers to the dataset with 58,000 molecules.

| Search algorithm | Unique new evaluations | Unique molecules with $F_{comb} > 0$ | Rate of discovery of molecules with $F_{comb} > 0$ | Molecules found better than starting dataset. (outstanding molecules) |
|---|---|---|---|---|
| BO-Learned | 841 | 458 | 54% | 4 |
| BO-Mord | 1273 | 406 | 31% | 5 |
| BO-Prop | 1799 | 893 | 50% | 2 |
| SUEA | 1037 | 1004 | 86% | 1 |
| EA | 1637 | 544 | 32% | 7 |
| Rand | 1236 | 0 | 0% | 0 |

### 3.c. Computational resources needed to run the algorithms

In this part, we discuss the impact of the computational time needed to run the search algorithm on the number of molecules evaluated when using a fixed computational resource. In the two sets of runs presented above, the number of unique new molecules that have been evaluated using the different search approach is different (first column of table 1 and 2). Although the same computational resources are allocated to all the different runs, the difference is caused by three aspects; 1) Some search algorithms take more computational time to suggest a new element to evaluate. For example, *BO-Learned* needs to generate the learned representation for many molecules before choosing the one with the highest acquisition function. The cost of this optimisation has a significant impact here because the time needed to evaluate a molecule is comparable to the time it needs to optimise the acquisition function. Improving the algorithm used to optimise the acquisition function could reduce this computational cost. 2) The computational time to evaluate molecules can vary from 3 to 20 mins depending on the size of the molecule (Figure S29-S30 in the SI). 3) When two separate runs simultaneously suggest the same molecule to evaluate, the calculations are run twice. Whereas, if a molecule that has been previously calculated, it will not need to be recalculated. This issue mainly affects the *SUEA*, as the different runs have a higher chance of suggesting the same molecules to evaluate at the same time. Further details about the computation time can be found in the supporting information Section 8.

Additionally, it is important to note that both *BO-Learned* and *SUEA* depend on a pretrained model, which in this instance was trained on data generated before the search approach began. The process of generating and training this model increases the computational cost of these methods, potentially making them less appealing when there is no initial data available.

### 3.d. Discussion of algorithm performance and surrogate models

Above, we have investigated the performance of six different search algorithms in exploring the chemical space of donor molecules for OPV applications. Our findings indicate that surrogate models significantly enhance the search algorithms' ability to identify superior molecules. The effectiveness of these algorithms in finding molecules above a certain threshold is closely tied to the accuracy of the surrogate models. In essence, more accurate surrogate models are generally beneficial for the search process. This observation aligns with several other studies which emphasize the critical role of model fidelity in guiding molecular discovery. [61, 62]

Additionally, we observed that when searching the full chemical space, algorithms with the best surrogate functions (such as *BO-Learned* and *SUEA*) tend to find fewer exceptional molecules compared to searches using less accurate surrogate models (like *BO-Mord*) or heuristic-based searches (*EA*). This discrepancy is due to the surrogate models' limitations in predicting molecules outside their training distribution, which includes these exceptional molecules. Furthermore, the strong performance of *EA* in discovering outstanding molecules is not unique to our study; Tripp et al. demonstrated that *EA* can often outperform more complex machine learning methods.[65] Overcoming this limitation in surrogate models, specifically their reduced generalizability to chemical regions under-represented or absent in training data, could be done through combining different model-based searches with heuristic search such as *EA*.[66]

The failure of the BO based algorithms in finding outstanding molecules is also related to the challenge of accurate uncertainty prediction of molecular properties. Improved uncertainty prediction requires an adapted molecular representation for the target application which is used to compute the distances/similarity between the molecules. In this work, we investigated three different molecular representations, and found that learned representations can outperform expert-curated ones. Furthermore, we demonstrated that achieving strong performance on a benchmark specifically tailored to the task does not necessarily lead to improved identification of exceptional molecules across the entire chemical space.[33, 67]

In the context of choosing the best search algorithm for the application at hand, we recommend using a combination of a surrogate model-based approach (*BO-Learned* in this case) with a heuristic based approach (*EA*). This combination would reduce the impact of the bias introduced by the surrogate model or the molecular representation. Coupling this strategy with in-depth analysis of the suggested molecules can help guide the search toward promising regions of the chemical space. To our knowledge, such detailed chemical space analysis is not yet fully automated and would still require a 'human in the loop'.[68-70]

## 4. Analysing the suggested molecules

We have demonstrated the use of six different search approaches to explore the chemical space of donor molecules constructed from various building blocks. In this study, we employed an evaluation function that focuses exclusively on the electronic and optical properties of the molecules, which can be computed relatively quickly using XTB and XTB-sTDA. This choice represents a compromise between relevance and computational efficiency. While more advanced evaluation functions are available within the same package, their use was beyond the scope of this work. Consequently, the molecules identified by the different search approaches can be considered as preliminary candidates for more detailed investigations.

The $F_{comb}$ distribution of all the molecules calculated over all the runs here (78,000 molecules) is significantly different to that of the initial benchmark dataset (Figure 6 c, d, where the grey shadow shows the distribution in the benchmark dataset). In the benchmark dataset, less than 1% of the molecules had $F_{comb}$ higher than zero, in the final dataset, more than 22% of the molecules did. This result confirms the performance of the different search algorithms compared to a random search. We also calculated the properties for ten of the best performing molecules using DFT/TDDFT, which confirms that the identified molecules are promising for the targeted application (more details are in the supporting information section 9). We have shown that finding molecules with ideal optical and electronic properties that match the requirement established is not particularly hard given these molecules are not rare (at least within the property range predicted by the computational setup used here). The next step would be to build on the findings of this work to establish harder requirements such as the synthesizability of the molecules, the molecular packing, and other properties impacting the exciton lifetime and the charge carrier transport.[71-73]

In Figure 6c and 6d, we show the impact of the presence of particular building blocks from a specific cluster on the performance of the molecules. Here, we find that the presence of more than two building blocks from cluster 4 (with the 3 rings fused structures) in the molecule results in a higher $F_{comb}$, with a mean above zero. This explains their overwhelming presence in the new dataset, almost 30,000 of the new molecules are in this category. In cluster 4, we can find the benzodithiophene (BDT) structures. BDT and its derivatives are the donor units in most donor polymers that show high power conversion efficiency in OPV devices that use Y6 as the acceptor molecule (e.g. PM6, D18 [44]). This confirms that our computational approach agrees with the current experimental results and efforts in finding good donor molecules for Y6. Two examples of the best performing molecules are shown in Figure 5a. Considering only the fragments in cluster 4, we find that among the same cluster, four specific building blocks are better than the rest, these are shown in the Figure 6b. It is interesting that the BDT unit is not among the absolute best building blocks; for example, 4,4'-alkyl-cyclopenta[2,1-b:3,4-b′]dithiophene (CDT) was more common in the best performing molecules. Experimentally, the CDT unit was common in donor polymers which performed better with fullerene-based acceptors.[74] On the other hand, the presence of building blocks from cluster 0 more than once in the molecules results in an overall reduced $F_{comb}$.

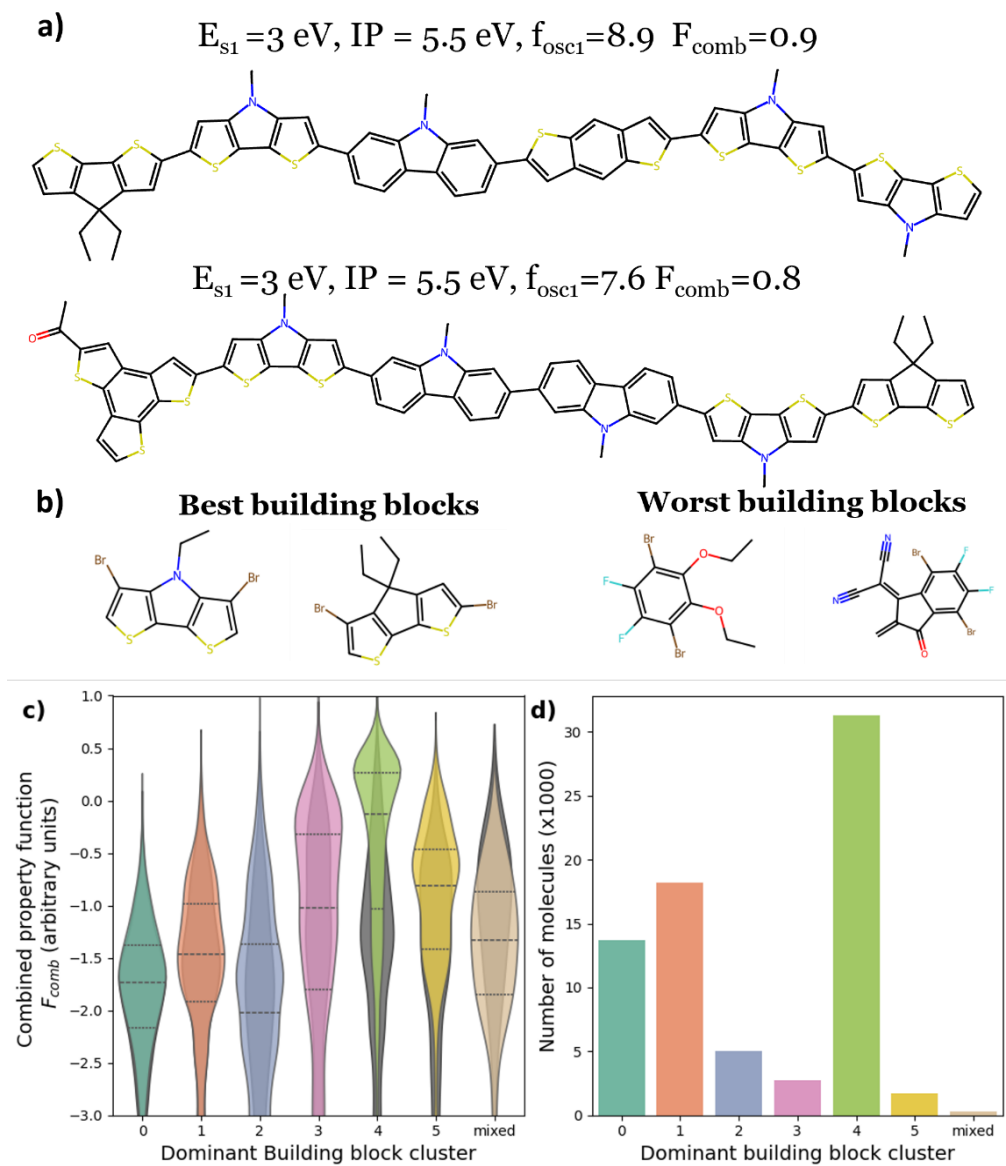

*Figure 6 Analysis of all oligomers predicted here across all runs. a) Examples of the best performing oligomers with their values of $E_{S1}$, IP and $f_{osc,1}$. b) The best and worst performing fragments based on their $F_{comb}$. c) Distribution of the combined property grouped by oligomers with building blocks from different clusters. Here we define the dominant building block cluster for a molecule as the cluster with the highest number of building block present in the molecule. If no cluster is dominant, we label the molecule as "mixed." The grey violin plots show the data in the benchmark dataset. The lines divide the distribution into quartiles. d) number of molecules in the dataset grouped by dominant building block cluster. The grey/darker bins show the data in the benchmark dataset.*

## Conclusions

We have introduced *stk-search*, a package to run search algorithms over molecules constructed from building blocks, and easily transferable to different use cases. The package considers the definition of the search space based upon the building block library provided, and the connectivity process for constructing molecules from the building blocks. The package also allows the use of different search algorithms including a) evolutionary algorithms which guide the exploration of the chemical space using rules similar to species evolution (*EA*), b) An enhanced version of the EA that uses a surrogate model to improve the selection of molecules to evaluate called Surrogate EA (*SUEA*), and c) an approach that uses the prediction of the molecules performance using a surrogate model as well as the uncertainty on this prediction, namely Bayesian Optimisation (BO). The package also offers different metrics to evaluate the performance of the search algorithms.

We used *stk-search* here to search the space of molecules for application as donors for organic photovoltaic (OPV) applications. We first assessed the overall performance of six different search approaches (including 3 different BO with different molecular representations) over a restricted precalculated benchmark space of 30,000 molecules. In these results we found a strong correlation between the search approaches that suggested overall better molecules considering the combined property function ($F_{comb}$) and their ability to find the molecules in the top 1% in the dataset. The explorations with *BO-Learned* and *SUEA* managed to find the best performing results molecules with the least number or iterations. The exploration with these search approaches found the highest number of molecules in the top 1% at the end of the 400 iterations.

When using the different search approaches over a larger search space of >$10^{14}$ molecules, we found that the performance differed significantly to that of the smaller search space. In most cases, the search approaches performed better than in the restricted space. The search algorithms using an efficient surrogate model (*BO-Learned*, *BO-Prop* and *SUEA*), showed a considerable increase in the rate of discovery of molecules with $F_{comb}$ above a specified target. On the other hand, the simple *EA*, or the *BO-Mord* using Mordred-based descriptors performed well in finding better molecules than the ones already in the dataset. The added complexity in defining a better representation or using a better surrogate model, only helped guide the search toward overall better performing molecules but fell short of finding molecules with properties better than the original dataset.

These results shed light on how we can use different search algorithms to explore the chemical space of molecules. We have also targeted the question of how we can assess different search algorithms before deploying them. Specifically, we have shown that testing the search algorithm in a benchmark dataset, however close the benchmark is to the task at hand, does not translate to a net improvement when deployed on a much larger chemical space. This discrepancy stems from fundamental differences between the small benchmark dataset and the full search space; most notably, the benchmark's unbalanced representation of the broader chemical space. Therefore, before the widespread use of a new and complex chemical space exploration method, we need to establish proper ways to evaluate them for the specific task. We suggest future work should focus on establishing metrics to evaluate the search space considered and improve the choice of a representative benchmark dataset to compare different search algorithms.

## Author contributions

Mohammed Azzouzi: Conceptualisation, Methodology, Software, Writing of first draft and manuscript, Funding Acquisition. Steven Bennet: Conceptualisation, Software, Draft Review. Victor Posligua: Conceptualisation, Software, Draft review. Roberto Bondesan: Methodology, Review, and editing. Martijn A. Zwijnenburg: Conceptualisation, Methodology, Review, and editing. Kim E. Jelfs: Conceptualisation, Methodology, Supervision, Funding Acquisition.

## Conflicts of interest

There are no conflicts to declare.

## Data availability

The code used in this study can be found in the GitHub repository: https://github.com/mohammedazzouzi15/STK_search.

The data generated in this work are stored in the materials cloud [https://archive.materialscloud.org/deposit/records/2338 ] in a format that can be added into a MongoDB database following a notebook provided.

## Acknowledgements

MA acknowledged the support of Eric and Wendy Schmidt AI in Science Postdoctoral Fellowship, a Schmidt Sciences program and the SNSF Swiss Postdoctoral Fellowship (grant no. TMPFP2_217256). K.E.J. thanks the Royal Society for a Royal Society University Research Fellowship and the European Research Council under FP7 (CoMMaD, ERC Grant No. 758370) and the Leverhulme Research Centre for Functional Materials Design for a Ph.D. studentship for S.B. MA and K.E.J acknowledge the support of the research computing service at imperial (Imperial College Research Computing Service, DOI: 10.14469/hpc/2232).

.

# *Supporting information*

## **Table of Contents:**

## **Computational details:**

### 1. **Search space definition.**

Our approach relies on first defining the search space through 1) the definition of the fragment library, 2) the way molecules can be connected, 3) restrictions on which fragment can be at different positions of the oligomer, as well as the overall symmetry for the oligomer. Our fragment library is a list of common fragments found in polymers used for optoelectronic applications.[5] We have limited the size of the fragments in this work to fragments with less than 30 non-hydrogen atoms. Then, we can define the building blocks by considering carbon atoms that could form single bonds with a neighbouring fragment. The carbon is then replaced with a "bromine" atom to generate building blocks for the *stk supramolecular toolkit* package. The current approach of defining building blocks can be further expanded by using the full capabilities of *stk* and considering different functional groups for the construction of the large molecules.[8]

### 2. **Evaluation function**

The next step in the search approach is to define how to evaluate the potential of certain molecule for a specific application. In the literature we can distinguish between property based evaluation function, which directly relate to relevant properties of the molecule for the target application,[9] and synthesis or accessibility based evaluation functions,[10] that focus on the synthesisability of the molecule and its ease of use for the application of interest. For example, in the case of organic electronic we are interested in how easy we can deposit this molecule on a surface to form a film.

In our case, we focus on property-based evaluation function as detailed in the main text. Following the construction of an oligomer using *stk*, we can use any quantum calculation package to calculate properties of the molecules. All the evaluations of the molecules were saved in a database that can be easily retrieved by the search algorithms, for this we used the *stk* capability of saving the constructed oligomers into a database in MongoDB.[75] Using *stk* we can generate InChIKey for the constructed molecules, that we can then use as a key to save the calculated properties of the molecules in separate databases. In this way, the evaluation method could be a multi-step process, and we can easily retrieve prior calculation outputs for other purposes.

Here, we use the capabilities of *stk* to build the oligomers and generate initial geometries for the quantum chemistry calculations packages. Then we optimise the geometry of the ground state using GFN2-xTB,[52] and calculate the ionisation potential and electron affinity using the IPEA option in xTB. The optical properties of the oligomers are calculated using sTDA-xTB.[53] Afterwards, the properties of the constructed molecules, as well as the xyz coordinates of the optimised geometry, are saved in the database.

3. **Search algorithm strategy.**

Starting from an initial population of molecules that will form our searched space, we use a search algorithm to suggest a new element in the search space that will be evaluated. The evaluated molecules will be added to the searched space, and we iterate the process for several iterations or until we reach the wall-time of our computational budget. We use a single molecule suggestion after each iteration; however the code can easily be adapted to run a batched search, i.e. where more than one molecule is suggested at each iteration.

When exploring unknown space, we can run several searches in parallel. The stochasticity of the search algorithms and the change in the initial population allows for a better exploration of the search space. Moreover, we use the database to avoid parallel runs re-evaluating the properties of the same molecules through a quantum chemical calculation, which is the time-consuming task in all the search approaches considered.

4. **Search algorithm Details:**

For our case, where we search over the space of 6-mer for OPV applications, we use the following details for the different search algorithms.
1) *Evolutionary algorithm (EA):* for an iteration we start from a searched space of molecules (current population), which are the molecules that have been evaluated so far during the current search. The searched space is formed of molecules considered in the initial population, and one molecule added after each iteration. We then form the parent population by selecting the top 3 molecules in the searched space and add 2 random ones from the searched space. Then we build the offspring population by performing all the mutation and crossover among the 5 constructed molecules in the parent population. Mutating a molecule in this case means we change a single building block in the molecules with one from the fragment database that satisfies the criteria used to define the search space. The crossover between two molecules means that we take the first $X$ building blocks of the first molecule and the last $N-X$ building blocks of the second molecule to build an offspring, $X$ here is a number from 1 to $N$ the total number of building blocks considered. From the offspring population generated, we select randomly a molecule. The selected molecules will then be evaluated using the defined evaluation function and added to the searched space. Compared to other evolutionary algorithms where we can define different selection algorithms to choose the molecules in the current generation to mutate or cross, we consider here a more straightforward method. We acknowledge that including different selection methods could potentially improve the search performance, however we wanted to show here the performance of the EA in its simplest form.

2) *Surrogate EA (SUEA):* we train the representation learning model described in the next section to learn the combined function values ($F_{comb}$). We use here a subset of the data we have already calculated in our database (for example either 10,000 molecules or 20,000 molecules). For the benchmark, we hide the top 1% of the oligomers from the training and validation dataset. We use the trained feed forward neural network to predict the combined function. The surrogate model is used to select the element in the offspring population generated using the same approach detailed in the EA section. We select the molecule with the highest predicted combined property among the offspring population for evaluation.

3) For the Bayesian optimisation, we use the RBF kernel with different length scale dimensions as implemented in GpyTorch.[40] We choose the RBF kernel following an initial trial where we tried different kernels such as the Mattern and the Tanimoto-based kernels. In that trial, the RBF kernel performed best, and we decided to use that here. We then use the *BoTorch* implementation to optimise the parameters of the kernel with a normalisation of the input data. Next, we use the Expected improvement (EI) acquisition function as implemented in *BoTorch*.[28] To optimise the acquisition function over the space, we considered the same implementation of the EA described above, apart from the fact that we consider a population of 1000 molecules from the offspring population to calculate the acquisition function at each iteration (compared to one evaluation every iteration for the global optimisation problem). We consider that the EA has converged in this case when the highest value of the acquisition function does not change over 5 iterations.

We implemented three different specific approaches that use BO:

a. Bayesian optimisation with representation from properties of fragment (*BO-Prop*). We use in this case, optical and electronic properties of the building block to generate a representation of the molecule. These properties are calculated using the same level of theory (xTB and sTDA-XTB). Specifically, we consider the number of atoms, the HOMO and LUMOs, the HOMO-LUMO Gap, the ionisation potential, the electronic affinity, the energy, and oscillator strength of the first 3 excited states. We have 12 properties for each building block which results in a 72-array representation for the 6-mers.

b. Bayesian optimisation with Mordred descriptors of the fragment (*BO-Mord*) here we consider for each building block the 1200 different 2D descriptors available in Mordred,[57] to generate a representation of the molecule. We also use principal component analysis to reduce the dimension to 100 which results in a representation of the constructed oligomer in the form of an array of 600 dimensions.

c. Bayesian optimisation with a deep Kernel (*BO-Learned*): in this case we represent the constructed molecules using the trained model (the same one used for the surrogate EA). We name this search approach deep kernel since we used a deep neural network to learn a numerical representation of the molecules, which subsequently will be used in the kernel to evaluate the covariance (similarity between molecules). The kernel can be expressed as:

$$K(CM_i, CM_j) = K\left(h(CM_i), h(CM_j)\right)$$

Where $K$ here is the RBF (radial basis function) Kernel, and $h$ represents the model used to learn the representation, $CM_i, CM_j$ are two constructed molecules.

5. **Model to learn molecular representations.**

In this subsection, we present the approach taken to generate molecular representations using deep learning models. This approach is considered to improve the correlation between the representation and the target property. This approach differs from the deterministic methods used in *BO-Mord* and *BO-Prop* in that the molecular representation is learned from a dataset where the target property serves as the label.

We use a deep learning model to learn the representation of the constructed molecules that is most correlated to the properties of interest. We use geometric models such as SchNet to relate the molecules represented as a point cloud with its XYZ coordinates and atomic types to an array representation of the constructed molecules [12, 15]. The geometric models considered use a graph representation of the molecules and use a message passing architecture to embed information's about the environment onto the graph nodes (atoms in this case). Different level of interactions can be considered, and this is one of the characteristics that distinguishes different models. Here, we will only present the use of one model (SchNet), but *stk-search* has other models included such as PaINN or SphereNet, following the implementation by Liu *et al.* [15].

We also included an additional neural network to predict the representation for the optimised xTB geometry of the molecule from the representation of the initial geometry. A detailed representation of the model structure is shown in Figure S1. During the training of the model, we generate the initial array representation of the constructed molecules, using

$$\alpha_{init} = h(CM_{init}) = h(f(BB_i))$$

where $h$ is the geometry representation model, $CM_{init}$ is the constructed molecules in the initial geometry. $BB_i$ is the i$^{th}$ building block, and $f$ is the function used in *stk* to construct the initial molecule. We then use the same method to generate the representation of the constructed molecule using the optimised geometry from xTB as

$$\alpha_{opt} = h(CM_{opt}) = h(f_{xtb,opt}(BB_i)).$$

We use the same geometry representation model to encode the representation of both inputs into an array of 128 dimensions, an array of this size showed the best performance in predicting the property following an initial hyperparameter search. Then we used a feed forward neural network ($FFNN_1$)

$$\alpha_{opt} \sim FFNN_1(\alpha_{init})$$

to transform the array representation of the constructed molecules using the initial geometry to the representation using the optimised geometry. This new representation is then passed through a feed forward neural network to predict the property of interest.

$$Target_{prop} \sim FFNN_2(FFNN_1(\alpha_{init})).$$

The loss function considered to train the model and update the weights and biases is a sum of the mean squared difference (*MSE*) between the two constructed molecules array representation and the *MSE* between the real and predicted target function.

$$\mathcal{L} = MSE(\alpha_{init}, \alpha_{opt}) + MSE(Target_{prop}, FFNN_2(FFNN_1(\alpha_{init})))$$

During the evaluation, we would only use the building block to generate the initial constructed molecules representation. This representation can either be used in conjunction with a simple kernel with a Gaussian process for the Bayesian optimisation approach. Or we can use the full model to predict the property of interest and use it as a surrogate model with the *SUEA*.

The choice of the loss function is critical as it directly influences the model's ability to learn and generalize from the data. The *MSE* is a common choice due to its simplicity and effectiveness in minimizing the error between predicted and actual values. However, it is important to consider the potential biases introduced by this choice. For instance, *MSE* assumes that errors are uniformly distributed, which might not be the case in real-world data. Alternative loss functions, such as contrastive loss or weighted loss functions, can be employed to address issues like class imbalance and improve model robustness. In this work, the learning curves with *MSE* were reasonably good not to require the use of other loss functions. We acknowledge however that further improvement on the performance of the surrogate model could be achieved through a different choice of loss function or better optimised learning scheduler.[76]

Computational details: the model was implemented in PyTorch using torch geometric for the graph neural network modules.

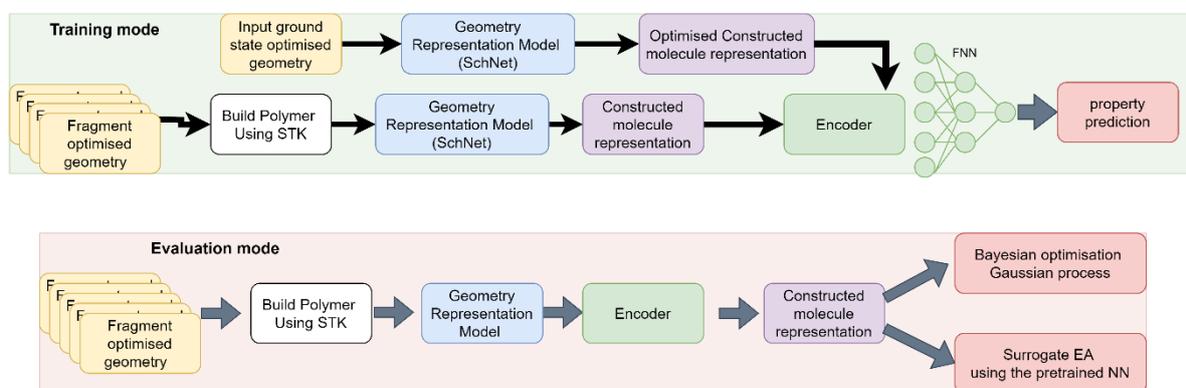

*Figure S1 Diagram representing the architecture of the model used to generate the oligomers representation.*

## The fragment database.

The library of building blocks we included are from fragments used in the organic solar cell community[5]. Here, we limit the building blocks to 30 non-hydrogen atoms. To generate a building block from a fragment, we replace a hydrogen atom with a bromine atom to define the connection points.

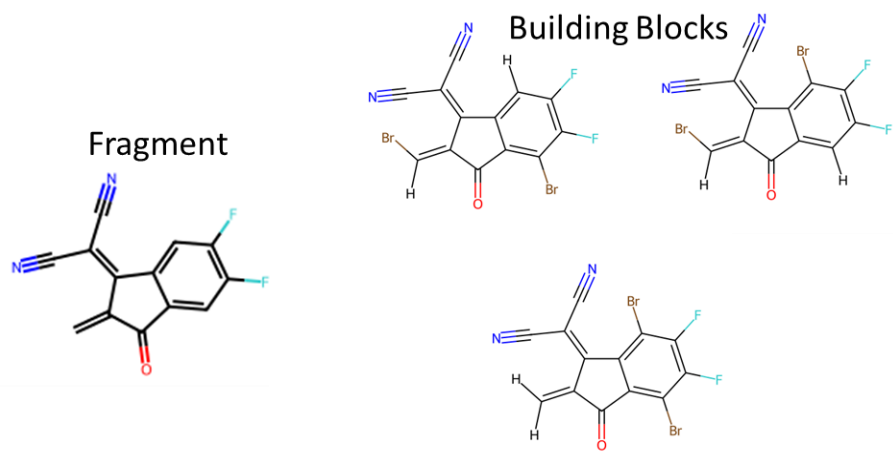

*Figure S2: Example of generating a building block from a fragment. Starting from a molecular fragment as on the left, we can define where the two-connection point with other building blocks by replacing hydrogen atoms with a bromine atom.*

The list of fragments is clustered using a mixture of manual and automated clustering using the Tanimoto similarity measure based on the ECFP fingerprints.[77] Here we restricted the number of clusters to 6 for ease of interpretation. The list of fragments in each cluster is shown the Figure below. Figure S3 shows the different cluster plotted in 2-dimensional space using the principal component analysis of the Tanimoto distance matrix between the different fragments.

# **Fragment list per cluster:**

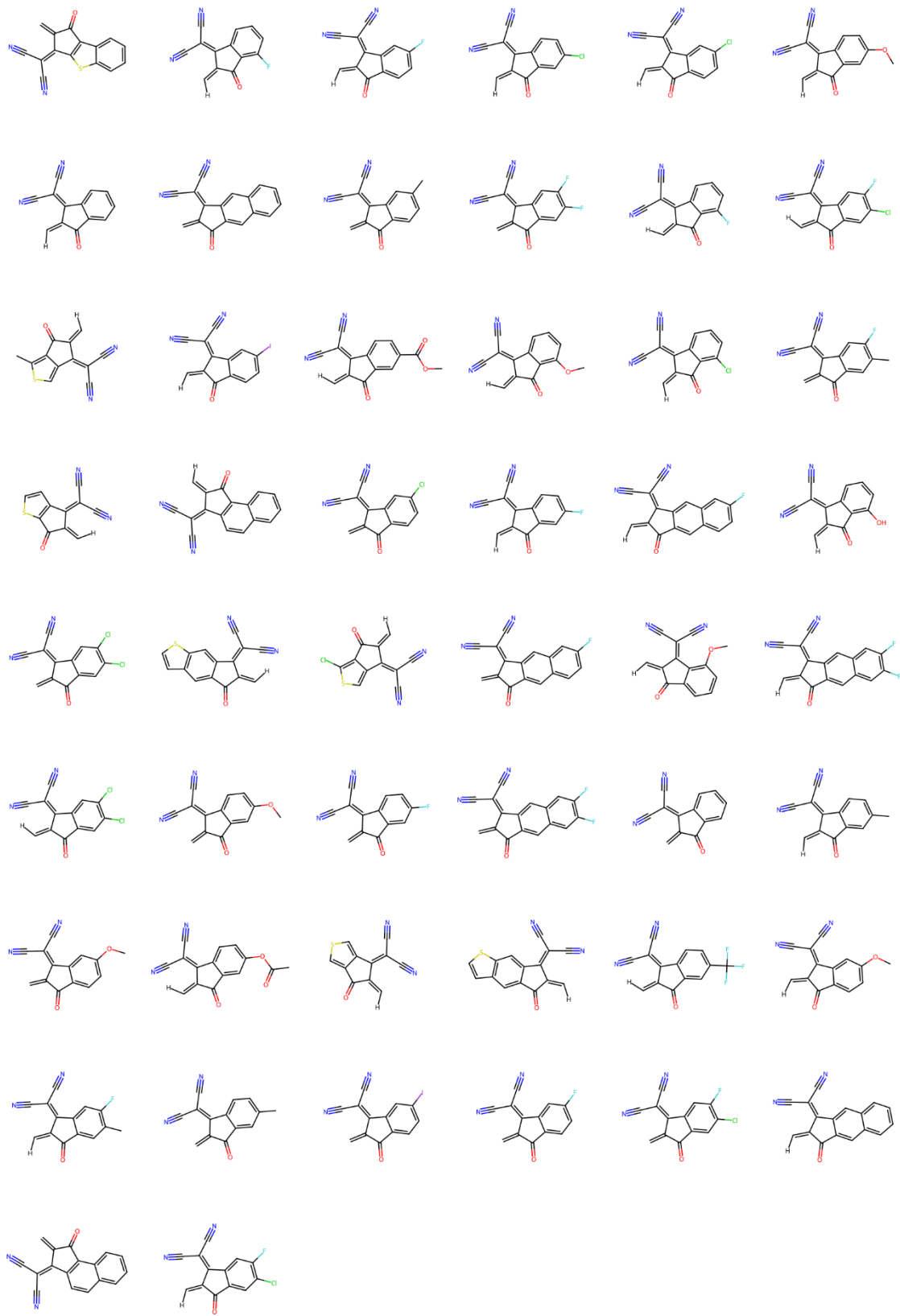

*Figure S3 fragment in cluster 0.*

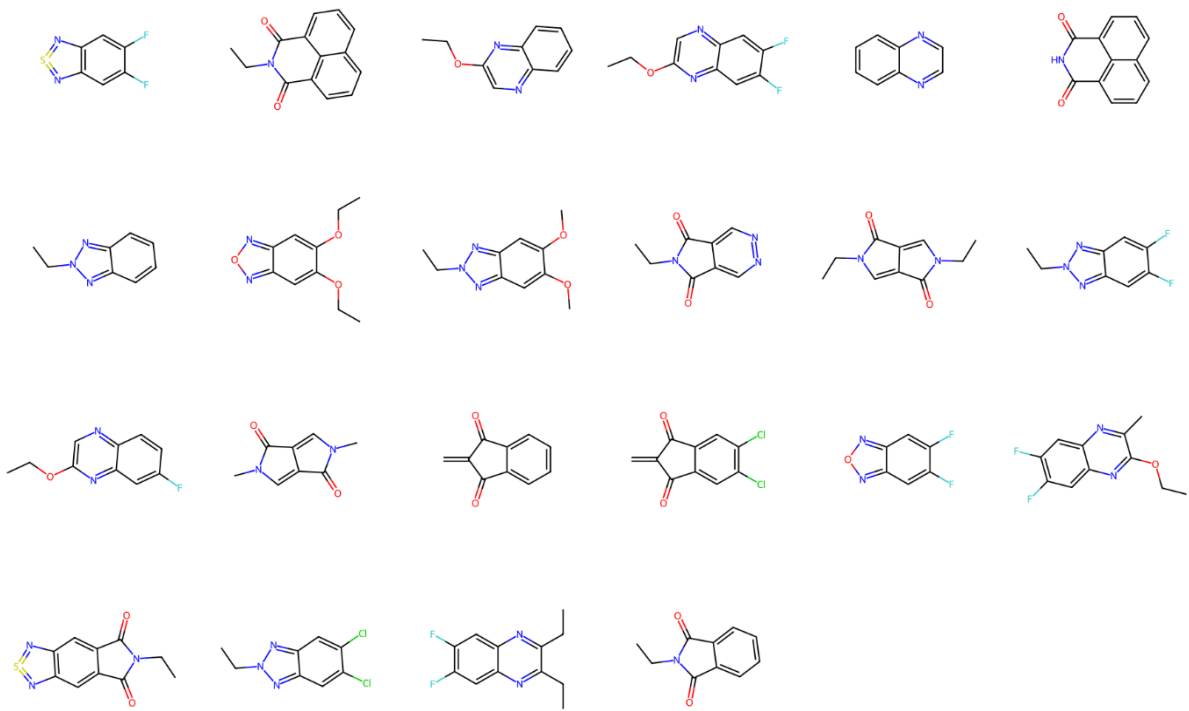

*Figure S 4 Fragment in cluster 1.*

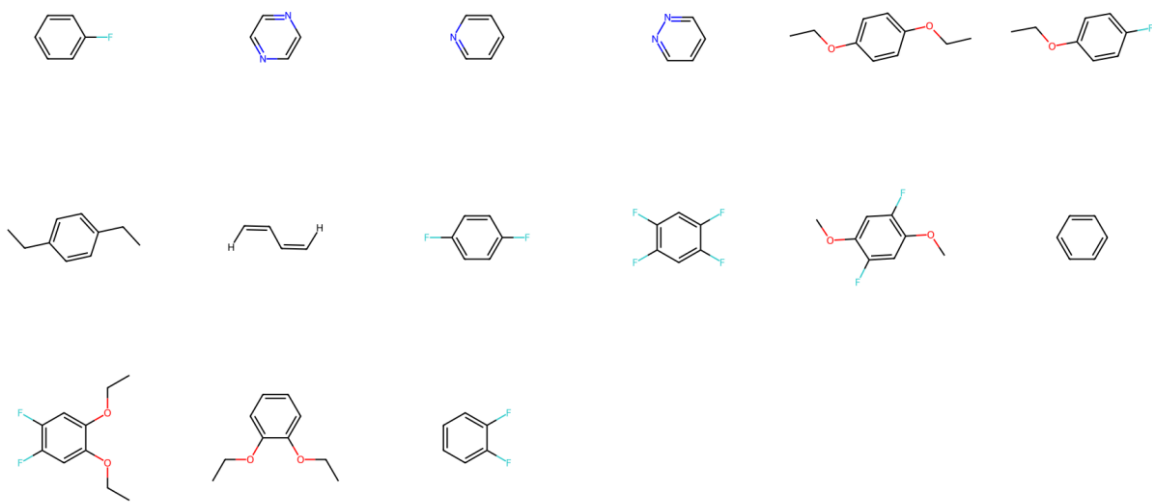

*Figure S5 Fragments in cluster 2.*

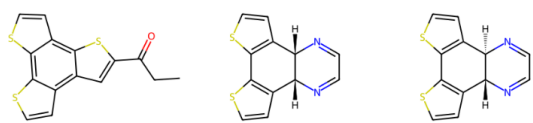

*Figure S6 fragment in Cluster 3*

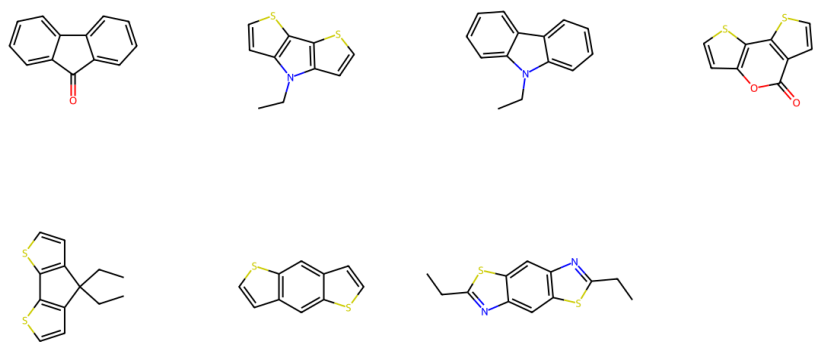

Figure S7 Fragments in Cluster 4

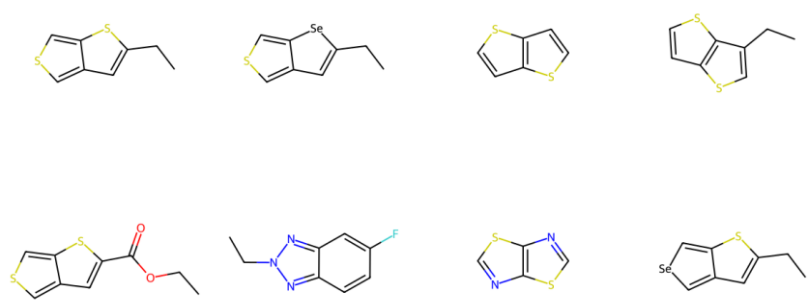

Figure S8 Fragments in Cluster 5

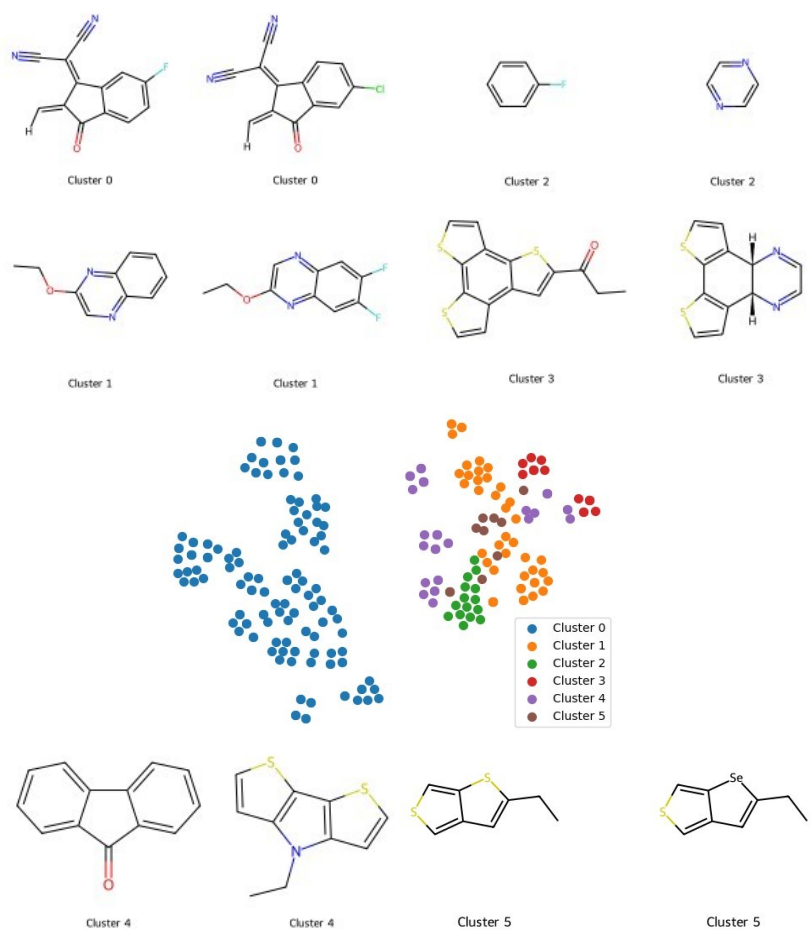

*Figure S9 Building block chemical space with representative structures from each cluster. Here the 2D representation of the chemical space is build using the Tanimoto distance matrix between the fragment represented using the ECFP fingerprint.*

Figure S10 shows the number of building blocks in each cluster considered. Here building blocks from cluster 0 are dominating the list. This is because there are many different versions of similar structure with fluorine or chlorine replacing the hydrogen atoms.

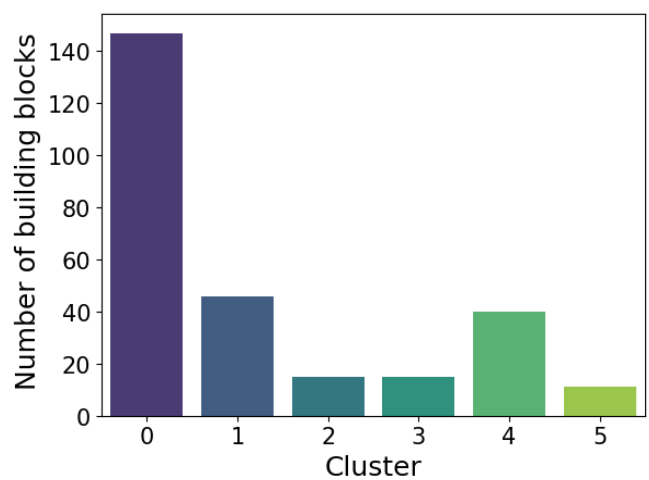

*Figure S10 Number of building block in each cluster.*

# Benchmark oligomer database

The molecules present in the benchmark dataset of 30,000 oligomers were randomly chosen from the >$10^{14}$ possible oligomers in the dataset. Figure S5 shows the number of oligomers with building blocks in different positions in the 6-mer. The distribution here is not representative of the proportion of building blocks in each cluster of the building block library (Figure S4). However, when considering the dominant cluster in each oligomer (Figure 2d), the distribution appears to be more representative. This difference can be explained by the very small portion of the full dataset considered in the benchmark (less than 1 in a billion). Hence not all the statistics of the original space are preserved in this benchmark dataset.

The distribution of the different oligomers properties of interest are shown in Figure S7. In this dataset, we find that the ionisation potential is centred around 6.5 eV, and the first excited state energy around 2.8 eV. The oscillator strength is overall low for most of the oligomers, with exceptions having an $f_{osc,1}$ higher than 2. For the combined property, only around 300 different oligomers (around 1%) have a combined property value higher than 0.

Figure S8, shows the distribution of the combined target, by building block in a cluster in the different position of the 6-mer. Here the difference in the distribution does not point toward a specific building block cluster that is considerably better or worse than the others.

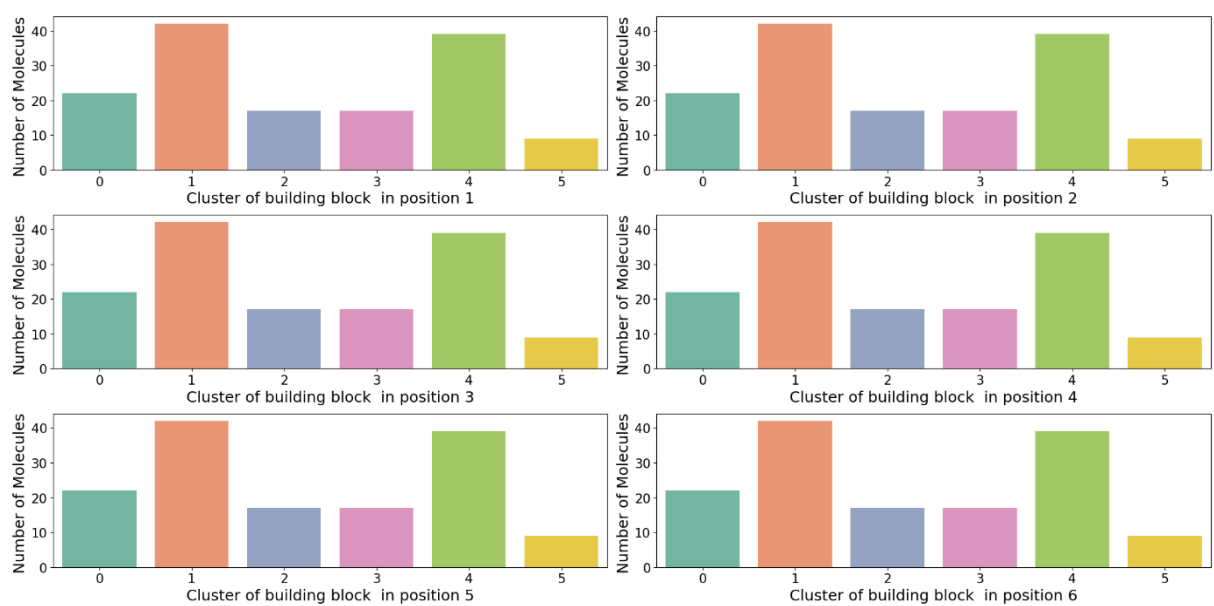

*Figure S11 Number of oligomers with building blocks from each cluster in oligomers in the benchmark dataset.*

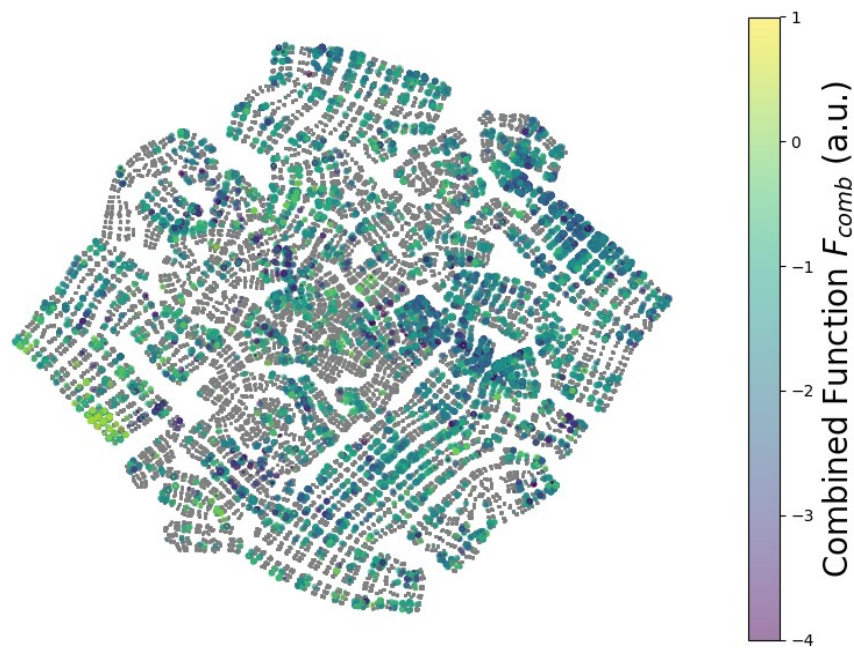

*Figure S12. A 2D representation of the chemical space based on a dimensionality reduction of a vectorial representation of the oligomers based where the fragments are represented by their numbered target values. The grey points show the extent of the full space here represented as $6^6$ unique points, (the number of fragment clusters is 6, and the total number of unique points would be $6^6$.) We used a TSNE dimensionality reduction to produce the 2D representation from the concatenated cluster array. The molecules in the benchmark dataset are here coloured by their combined property values according to the scale on the right of the figure.*

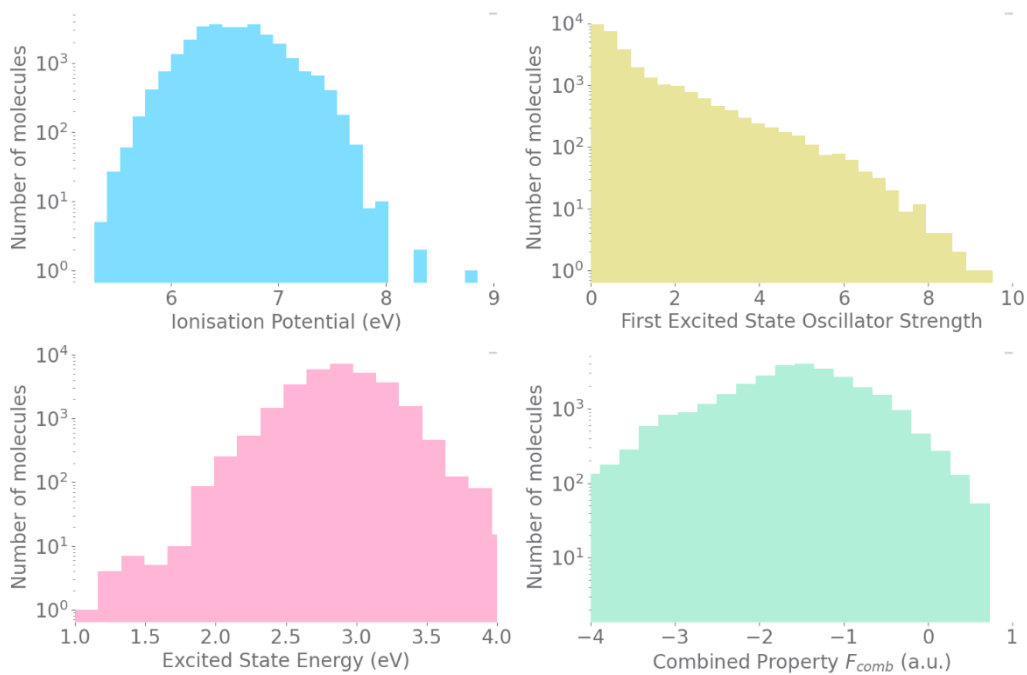

*Figure S13 Distribution of the different oligomer predicted properties in the benchmark dataset.*

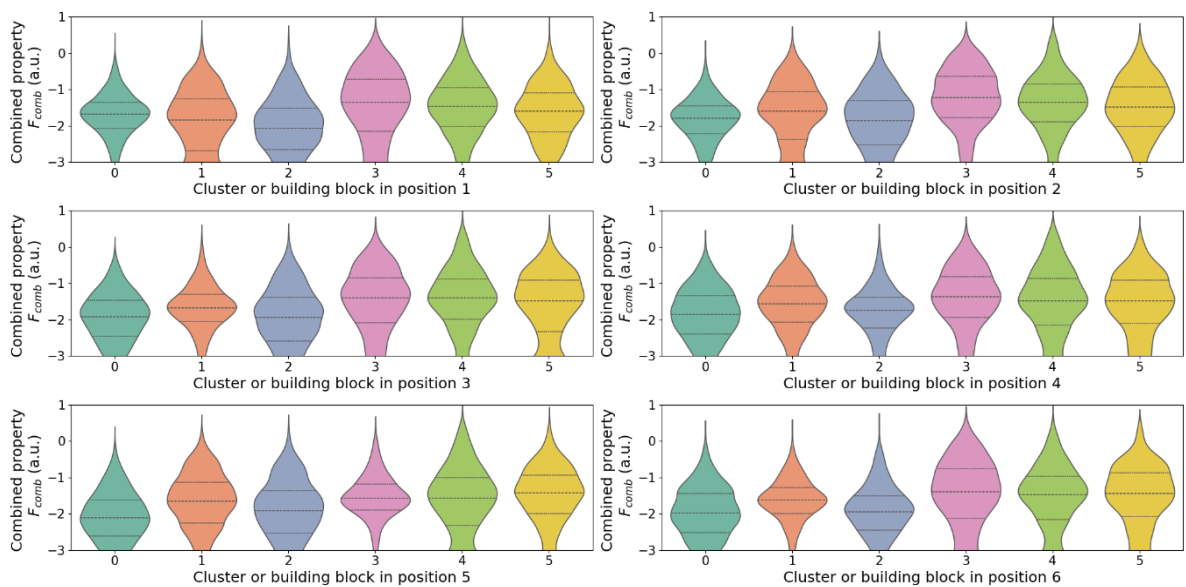

*Figure S14 The oligomers' combined property ($F_{comb}$) distribution with building blocks from different clusters in different positions. The different positions here refer to the position of the building block in the oligomer chain. This Figure shows the distribution of the combined property with the presence of building blocks from the different cluster can impact the distribution of the combined property. For example, in this dataset molecules with building block from cluster 3 in the first position (position 0) have a higher overall combined property (pink violon plot).*

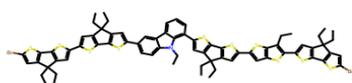 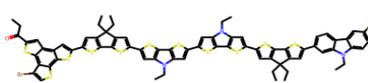 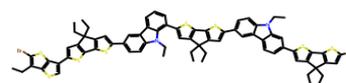

DC 4, CP 0.73, ES1 2.61,IP 5.48, fosc1 8.78    DC 4, CP 0.71, ES1 2.72,IP 5.48, fosc1 7.45    DC 4, CP 0.71, ES1 3.03,IP 5.54, fosc1 5.82

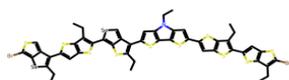 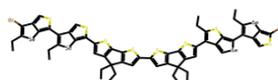 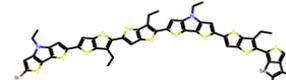

DC 5, CP 0.54, ES1 2.69,IP 5.66, fosc1 7.14    DC 5, CP 0.39, ES1 2.93,IP 5.81, fosc1 5.50    DC 5, CP 0.25, ES1 2.61,IP 5.68, fosc1 4.21

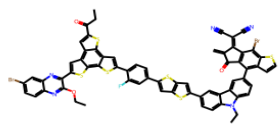 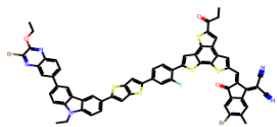 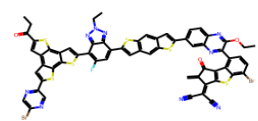

DC mixed, CP -0.07, ES1 3.09,IP 6.23, fosc1 5.03    DC mixed, CP -0.20, ES1 2.95,IP 6.30, fosc1 4.27    DC mixed, CP -0.26, ES1 2.80,IP 6.24, fosc1 3.79

*Figure S 15 Example of best molecules with dominant clusters from cluster 4, 5 and the mixed case. Here DC stands for dominant cluster of building blocks. The dominant cluster is the building block cluster with the highest number of building block in the molecules. Mixed here refer to the case where there is no dominant cluster.*

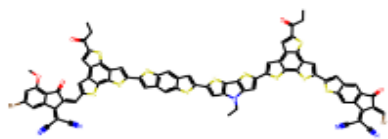
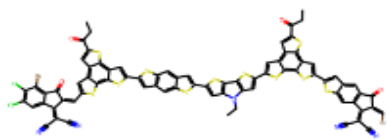
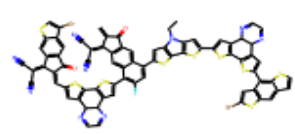

DC 0, CP 0.01, ES1 2.73,IP 6.21, fosc1 7.01    DC 0, CP -0.01, ES1 2.71,IP 6.26, fosc1 7.77    DC 0, CP -0.01, ES1 2.64,IP 6.01, fosc1 4.70

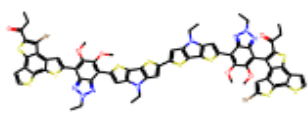
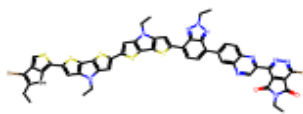
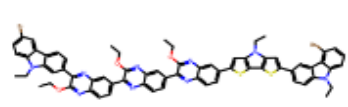

DC 1, CP 0.52, ES1 2.71,IP 5.62, fosc1 6.05    DC 1, CP 0.42, ES1 3.25,IP 5.67, fosc1 5.19    DC 1, CP 0.39, ES1 2.79,IP 5.79, fosc1 6.09

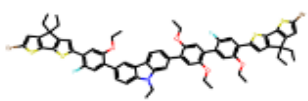
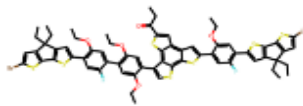
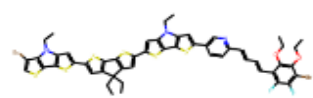

DC 2, CP 0.29, ES1 2.69,IP 5.79, fosc1 5.38    DC 2, CP 0.27, ES1 3.18,IP 5.76, fosc1 4.19    DC 2, CP 0.25, ES1 3.02,IP 5.89, fosc1 4.47

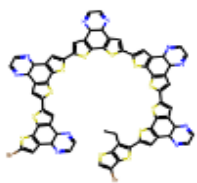
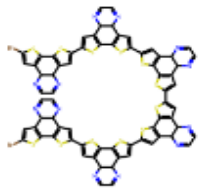
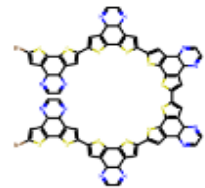

DC 3, CP 0.59, ES1 2.67,IP 5.52, fosc1 5.94    DC 3, CP 0.59, ES1 2.67,IP 5.52, fosc1 5.94    DC 3, CP 0.58, ES1 2.71,IP 5.59, fosc1 6.55

*Figure S 16 Example of best performing molecules with dominant building blocks cluster 0,1,2,3. Here DC stands for dominant cluster of building blocks. The dominant cluster is the building block cluster with the highest number of building block in the molecules. Mixed here refer to the case where there is no dominant cluster.*

# Performance of the search algorithms on the benchmark dataset

## 1. Performance after 400 iterations

To assess the performance of the search algorithms, we introduce a few extra metrics related to the similarity between the molecules suggested during the search. We distinguish between:

1) Similarity of the suggested oligomers to the initial population. This is calculated as the maximum Tanimoto similarity between the oligomers suggested and the oligomers in the initial population [63]. This metric helps assess how the outcome of the search algorithm depends on the initial starting point, and how the initial population impacts the way a search algorithm explores the chemical space.

2) Similarity distribution of the oligomers suggested. This is calculated as the mean similarity between the oligomers suggested by the search algorithm (i.e. considering all the molecules suggested in the previous iterations). This metric helps assess how the search algorithm explore the space. A high value of this metric means that the search algorithm focuses on a specific area in the chemical space.

The assessment of the 6 different search approaches using the two metrics above is shown in Figure S18. First, we consider the similarity of the molecules suggested to those considered in the initial population. Here we find that the *EA* seems to be the most bound to the initial population considered, where the *BO-Learned* shows similar behaviour to the random search. It is worth noting that here we consider the Tanimoto similarity based on the ECFP fingerprint, which are considerable different to the representation considered in the *BO-Learned*. For the similarity between the oligomers suggested by the search algorithms (Figure S10b), the *BO-Learned* and *SU-EA*, suggest very similar molecules after the first few iterations. Due to the more exploratory nature of the *BO-Learned*, it starts suggesting more diverse oligomers after 100 iterations.

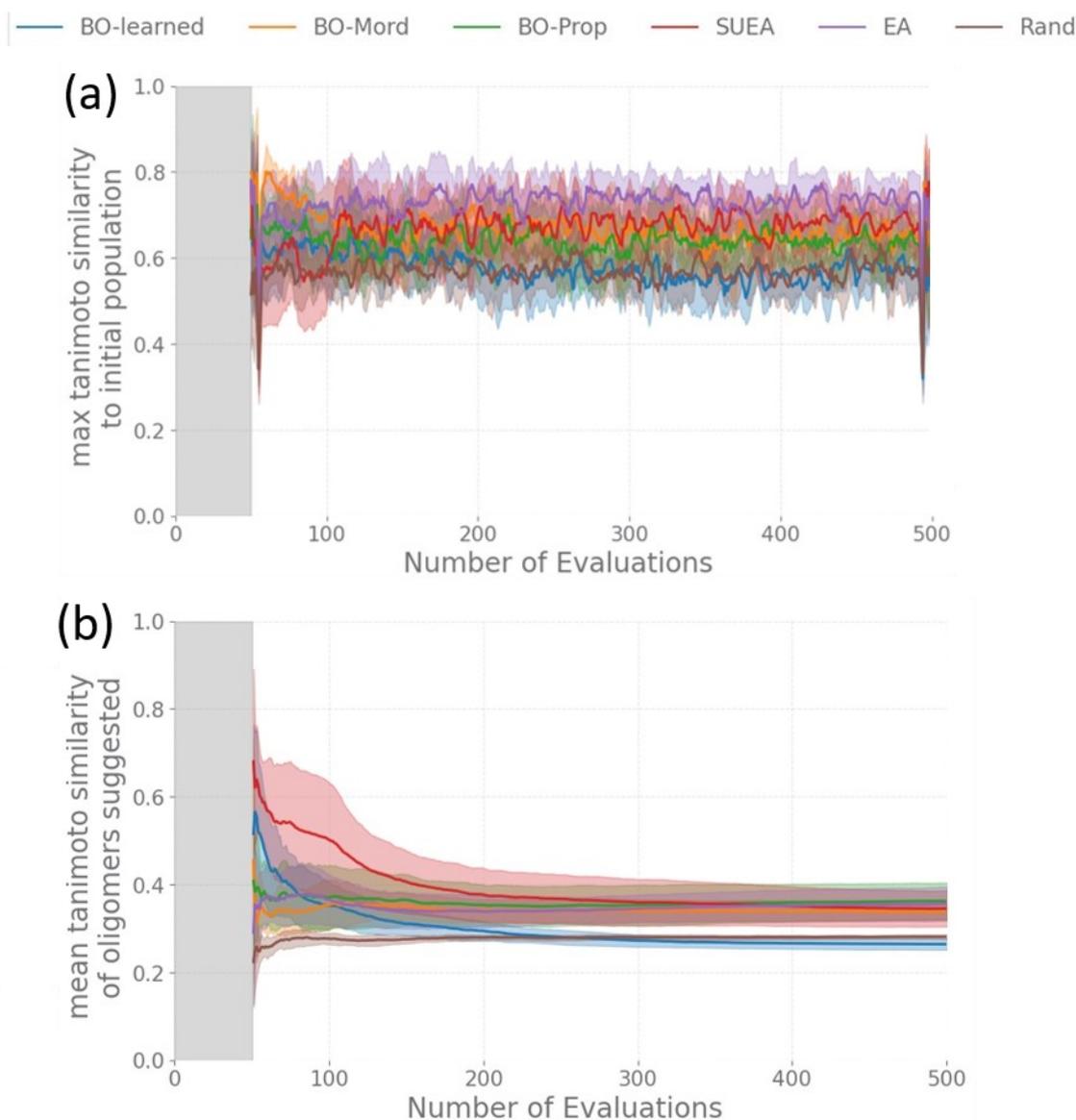

*Figure S17 a) maximum Tanimoto similarity of the molecules suggested at the current iteration to the elements in the initial populations of molecules. b) average Tanimoto similarity between the molecules suggested by the search algorithm. The grey shaded area shows the first 50 initial iterations. The colour shaded area around the curves shows the deviation of the metric within 1 variance. The variance is calculated here over 25 independent runs with different initial populations.*

In Figure S19, we show a 2D representation of the search space using the different representations considered for the BOs. The figure shows how the learned representation helps smoothing the chemical space and brings oligomers with similar target property closer. The choice of the representation can make the search space to explore smoother and easier to explore, grouping the best performing oligomer in space. Hence the Gaussian process used in the BO will perform significantly better at predicting the performance of the oligomers in the search space. In orange scale, we show in Figure 5, the elements suggested by a *BO-Learned* search. In the space of learned representations, we can see that the search algorithm mostly suggests elements that are close to the best ones seen in the dataset. Then it starts exploring the space more where it has higher uncertainties. The learned representations are however not perfect for the top 1000 oligomers not considered when learning the representation, as shown by the performance of the predictive model on those data.

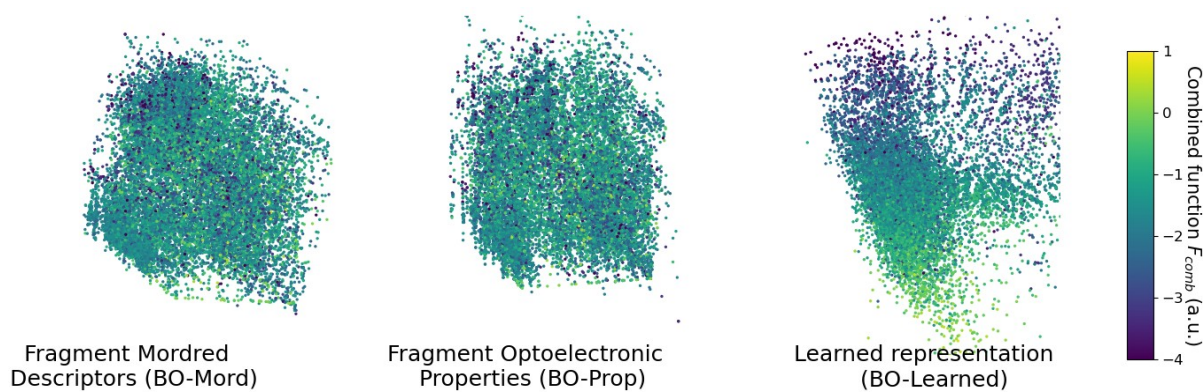

*Figure S18 A 2D representation of the chemical space of precalculated oligomers. The representation is based on a PCA of the used representations for the different BO search algorithm. The evolution of a search algorithm using BO with the learned representation and an expected improvement acquisition function is shown in orange scale. The scale goes from light to dark with the number of iterations. The other colour scale shows the combined target value.*

In Figure S20 and S21, we look at the performance of the search algorithm considered across the set of 25 runs. Here we consider the distribution of the combined property for the unique oligomers suggested by each algorithm. The oligomers suggested by the random search have a similar distribution to the oligomers in the benchmark (shaded histogram in the plots). In this limited search space, we find that the *BO* and *SUEA* search algorithms, suggest overall elements with higher target property than the random search. It is hard to distinguish the different BOs in these graphs, as they all show similar distributions. The *EA* shows the least difference with the *Random search*, which explains its limited performance in this search space.

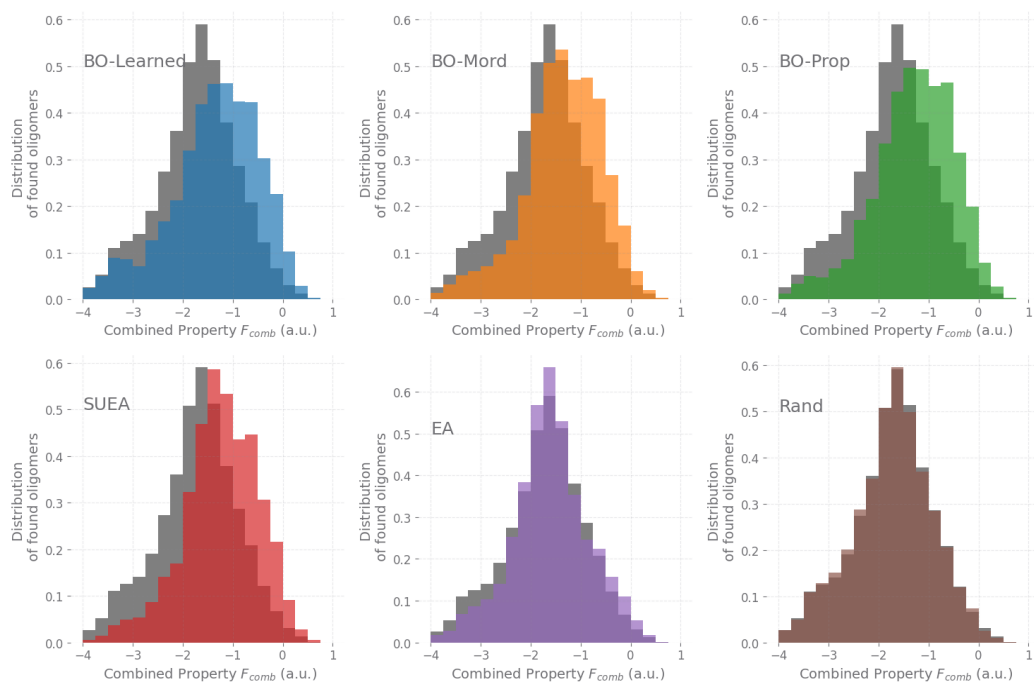

*Figure S19 Distribution of the combined property of the unique oligomers suggested by the different search algorithm over 25 independent runs with different initial populations. The black shaded histogram shows the distribution of the combined property of the oligomers in the benchmark.*

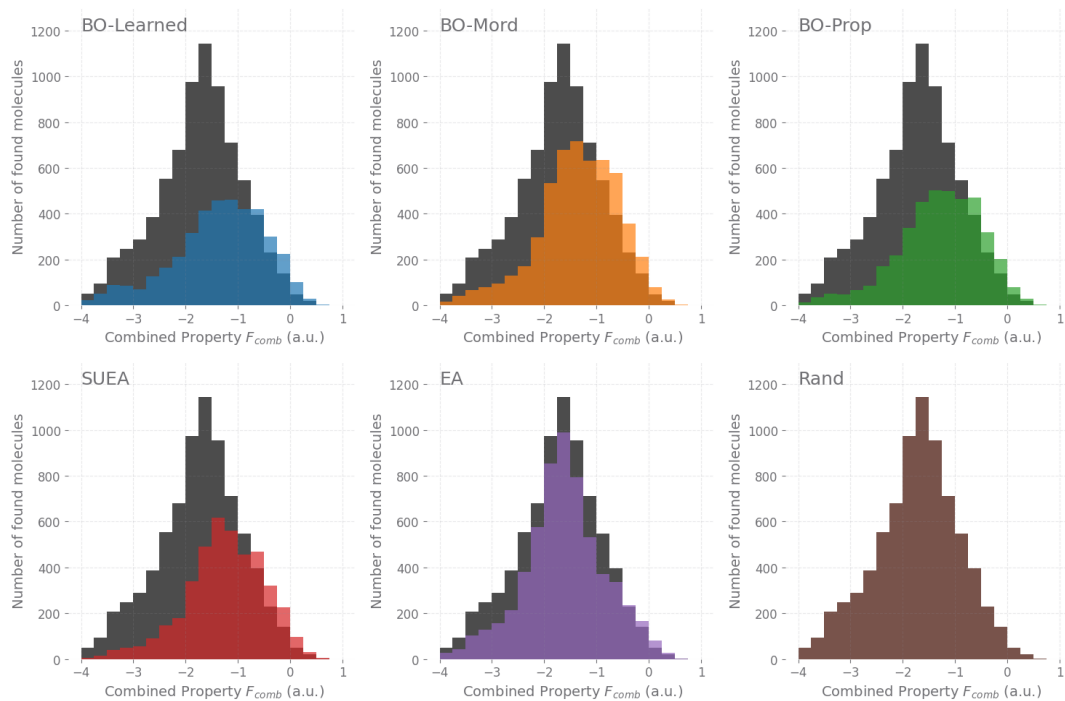

*Figure S20 The number of oligomers found with target property with the different search algorithms over the benchmark search space. The panels show the distribution of the unique oligomers suggested by the search algorithms over 25 independent runs. The black shaded histogram shows the distribution of the randomly suggested elements to evaluate to compare with the different search algorithm.*

## 2. Performance after 800 iterations

In order to investigate how the performance of the different algorithm changes when we run the search for a larger number of iterations, we have rerun the different search algorithms for 800 iterations. We limited the number to 800 for computational reasons as the BO based model require retraining a gaussian process that scales badly with the number of iterations. Figure S21 shows the performance of the search in terms of the best molecules found, the mean value of the target molecule suggested, as well as the number of top molecules found. In this case, the results are similar to those in figure 3, *BO-Learned* and *SUEA,* find the best molecules the fastest and overall suggest molecules with higher target property. We did not consider here the *BO-Mord* as the size of the representation increases the computational load considerably.

After 300 iterations, random performs better that EA in finding the molecules with the higher $F_{comb}$. This is related to the limited search space, where although the EA keeps predicting molecules with a higher mean $F_{comb}$, some of the higher performing are only accessible though mutation of very specific molecules. Hence the EA has higher chances of being stuck in an area of the chemical space, with less chances of overcoming it.

After 700 iterations, Rand is the best model at consistently finding molecules with the highest $F_{comb}$. Apart from the improved performance of Rand over higher number of iterations, the models perform similarly to the case in figure 3 of the paper. With increased number of iterations, the mean $F_{comb}$ of the molecules suggested decreases for BO-learned, BO-prop and SUEA; the rate of discovery of best molecules decreases as well and gets closer to the Rand case.

We further investigate why the search algorithm perform worse in finding the top molecules after a large number of iterations as compared to a random search. In our extended analysis of the EA and SUEA algorithms, we find that the discovery of top-performing molecules is strongly correlated with their mutational accessibility. In this context, mutational accessibility refers to how easily a target molecule can be reached from other molecules in the dataset through a series of small modifications—specifically, changes in their constituent building blocks. Within the limited chemical space of our benchmark, molecules that share at least four building blocks with a top-performing candidate are significantly more likely to guide the algorithm toward that candidate.(Figure S22 b)

For Bayesian Optimization (BO)-based models, we observe a different but related form of bias. These models are influenced by the unbalanced distribution of building blocks in the dataset. The surrogate model, trained on this skewed data, tends to favor regions of chemical space populated by frequently occurring building blocks. This bias is further reinforced by the Expected Improvement (EI) acquisition function, which prioritizes candidates that are structurally similar to many existing molecules.(figure S22 a) As a result, molecules composed of underrepresented building blocks are less likely to be selected, even if they have high potential. This is evidenced by the correlation between the likelihood of discovering top candidates and the number of molecules in the dataset that share at least four building blocks with them.

The results from running the search algorithms over 800 iterations further highlight the impact of dataset imbalance. While the benchmark dataset was generated randomly, it still exhibits structural biases common in chemical datasets. These biases increasingly affect algorithm performance as the search space coverage grows. To mitigate this issue, one could consider either (i) a balanced sampling strategy that ensures a more uniform distribution of building blocks, or (ii) a significant expansion of the benchmark dataset to dilute the effects of imbalance. However, the former approach may unfairly favor model-based algorithms over random search, while the latter would require substantially more computational resources and a redefinition of the study's scope.

Here we are focusing on the performance of the search algorithms on the limited search space of 30,000 molecules; and we allow for the exploration of more than 2% of the search space. Reaching a similar value on the full search space would mean exploring more than $10^{12}$ molecules, which would not be feasible. This explains the overall very low performance of Rand on the full search space (figure 4).

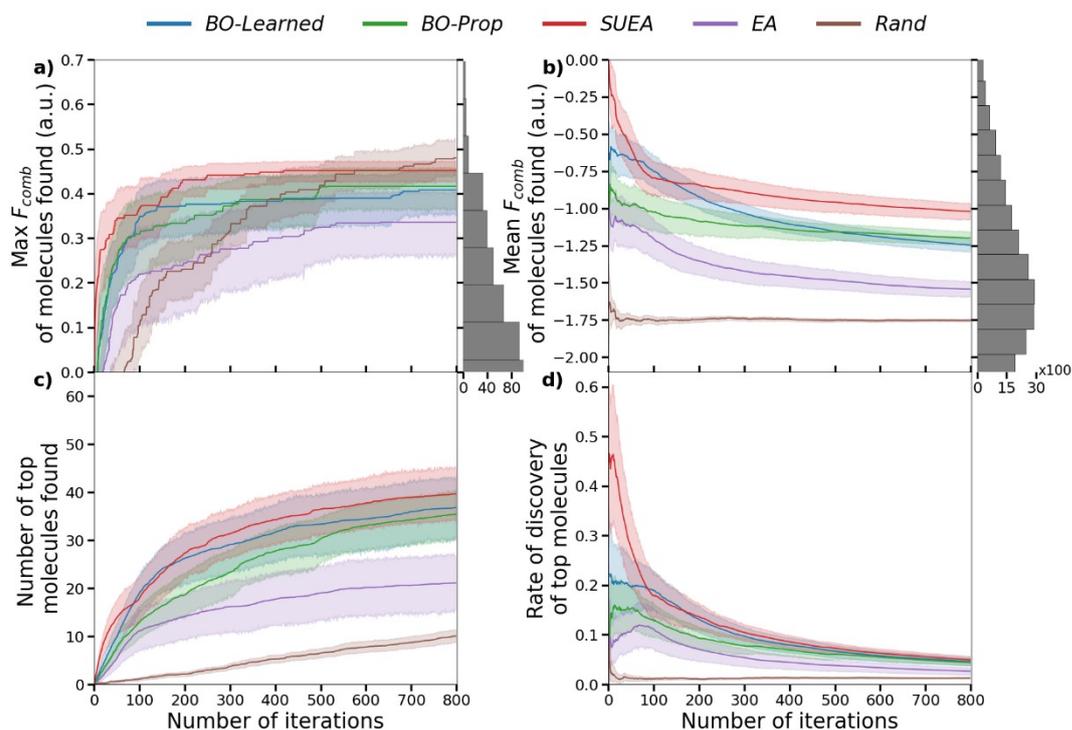

*Figure S 21 Performance of the six different search approaches on the precalculated benchmark dataset of 30,000 molecules. The solid-coloured lines show the mean $F_{comb}$ over 25 runs with different initial populations and the coloured shaded area shows the variance of the $F_{comb}$ over those different runs; a) Maximum $F_{comb}$ found for an oligomer up to the current iteration. The histogram on the right shows the distribution of the oligomers in the benchmark dataset; b) Mean $F_{comb}$ of the oligomers found up to the current iteration; c) Number of oligomers in the top 1% found up to the current iteration (top 1% is 300 molecules); d) Discovery rate of the top 1% oligomers in the dataset, calculated as the (number of top molecules found)/(number of iterations).*

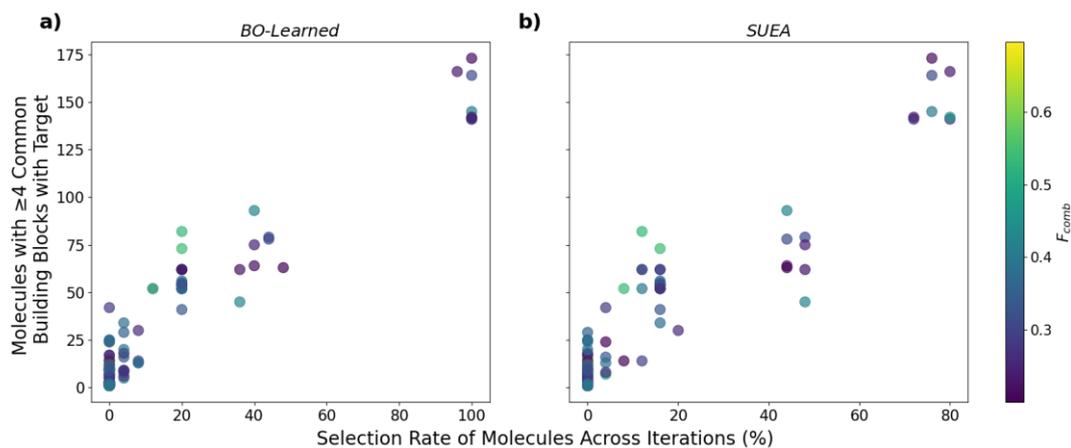

*Figure S 22: Correlation between the selection rate of top-performing molecules and the number of dataset molecules sharing four building blocks with each top molecule. The selection rate is averaged over 25 independent runs after 800 iterations. Results are shown for (a) BO-Learned and (b) SUEA.*

# Impact of THE training dataset on the performance of the *BO_Learned* algorithm on the benchmark dataset

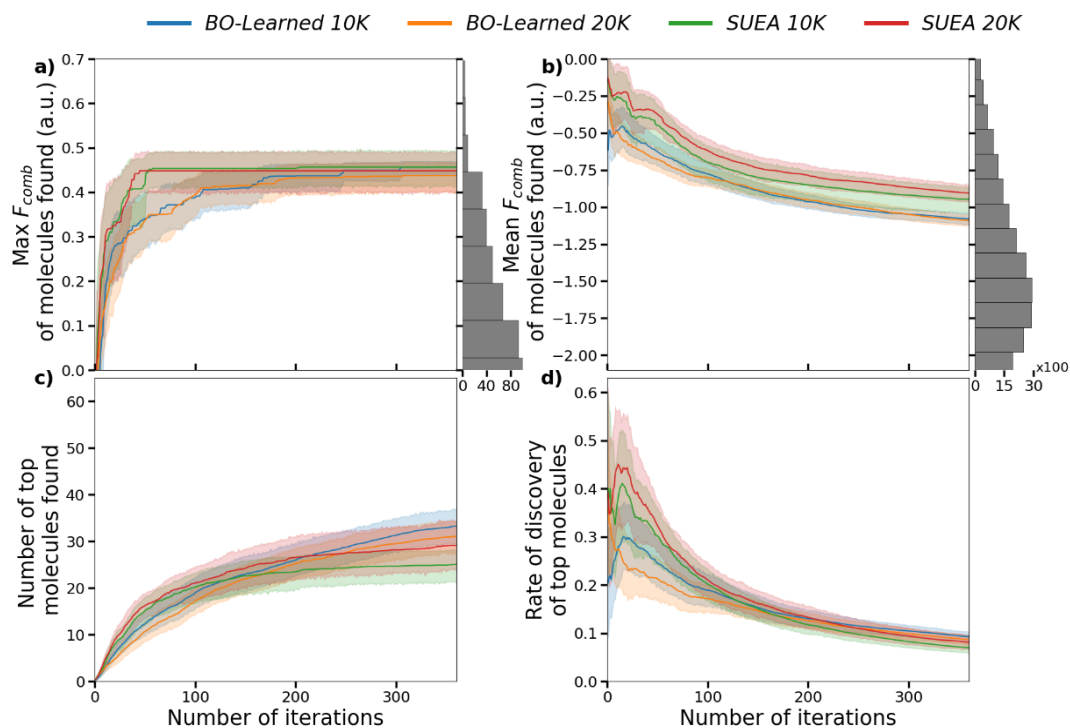

*Figure S 23 Impact of the number of datapoint in the training dataset on the search algorithm performance. The performance of the search algorithms is assessed when exploring the space of precalculated oligomers (~30000 unique oligomers). The solid-coloured lines show the mean $F_{comb}$ over 25 runs with different initial populations and the coloured shaded area shows the variance of the $F_{comb}$ over those different runs; a) Maximum $F_{comb}$ found in a molecule up to the current iteration. The histogram on the right shows the distribution of the molecules in the benchmark dataset; b) Mean $F_{comb}$ of the molecules found up to the current iteration; c) Number of molecules in the top 1% found up to the current iteration (top 1% is 300 molecules); d) Discovery rate of the top 1% molecules in the dataset, calculated as the (number of top molecules found)/(number of iterations).*

To assess the impact of the number of trainings datapoints, we assess the performance of the search algorithms (*SU-EA* and *BO-Learned*) with two sizes of training datasets (10k and 20k). First, we compare the performance of the models over the different dataset as shown in table S1 and Figure S24 and Figure S25. The performance of the model trained with 20,000 molecules on the test set is reduced compared to the one with 10,000. This can be related to the different training sets considered, where a different split of the dataset can result in a different performance. For the test dataset considered here, we include the top 300 molecules in the benchmark dataset, as we are interested in how the models would perform in finding the best molecules in the dataset. The split between the train and validation dataset is done randomly, with 90% of the molecules considered for the training and 10 % in the validation dataset.

The observed change in the performance of the surrogate model on the test data did not have an impact on the performance of the search algorithms (Figure S23). In this case, we find that the increase of the number of datapoints had a minor impact on the performance of the search algorithm which agrees with the minor improvement in the performance of the trained models (table S1). Apart from the quantity of the data, it is also important to use a dataset that is diverse and representative of the search space. We have not fully explored that avenue in this work, and it would be an interesting follow up study. In the case of the *BO-Learned*, the case with 20k datapoints performed overall better than the one with 10k datapoints. For the *SU-EA*, the trend is inversed, pointing toward the limited performance of the surrogate model.

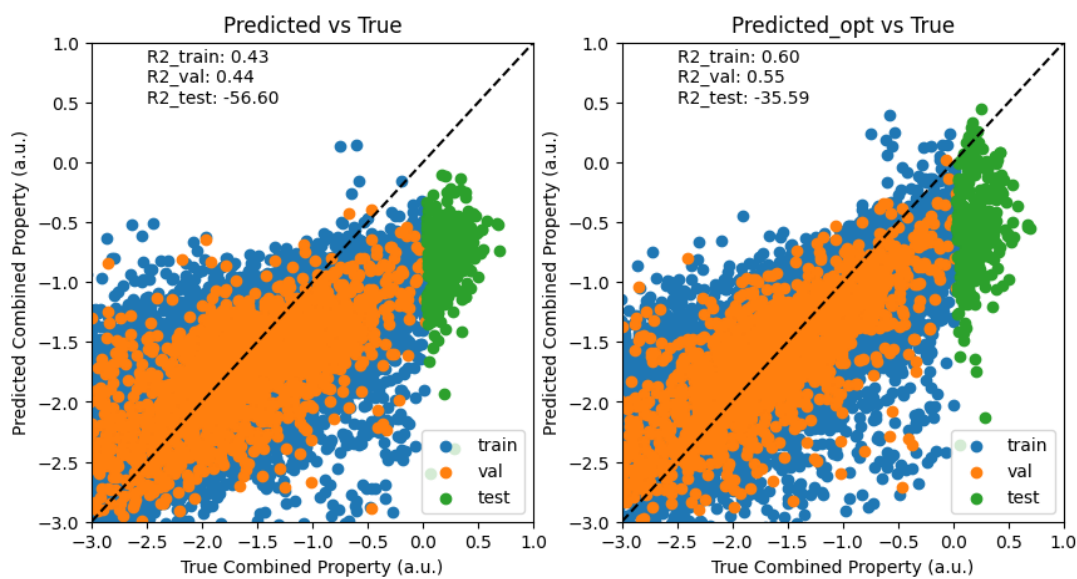

*Figure S24 Performance of the pretrained model with a training dataset size of 20,000 oligomers. the right panel shows the predicted combined property using the optimised geometry of the oligomers. The left panel shows the performance of the full model considering the geometry generated using stk alone. The performance drops as we try to learn a mapping between the initial geometry from stk to the optimised one.*

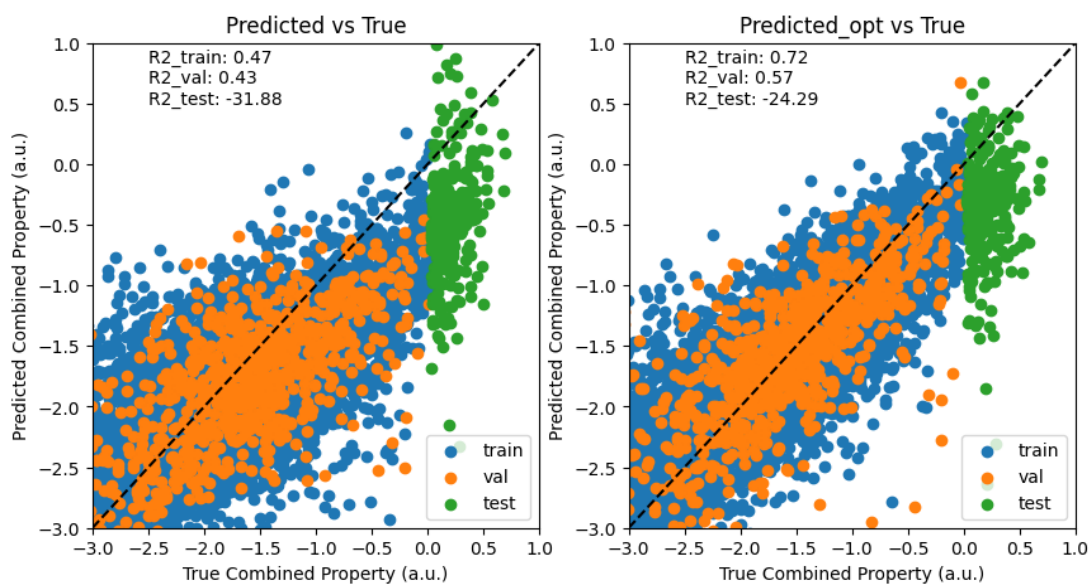

*Figure S25 Performance of the pretrained model with a training dataset size of 10,000 oligomers. The right panel shows the predicted combined property using the optimised geometry of the oligomers. The left panel shows the performance of the full model considering the geometry generated using stk alone. The performance drops as we try to learn a mapping between the initial geometry from stk to the optimised one.*

*Table S1 performance of the pretrained model with different training dataset size. Mean average error (MAE) is the metric used in this table.*

| Number of oligomers in training dataset | 10,000 | | | 20,000 | | |
|---|---|---|---|---|---|---|
| Metrics | MAE Train set | MAE Val set | MAE test set | MAE Train set | MAE Val set | MAE test set |
| Values | 0.46 | 0.48 | 0.71 | 0.47 | 0.45 | 1.02 |

# Impact of choosing different acquisition function on the search algorithm performance

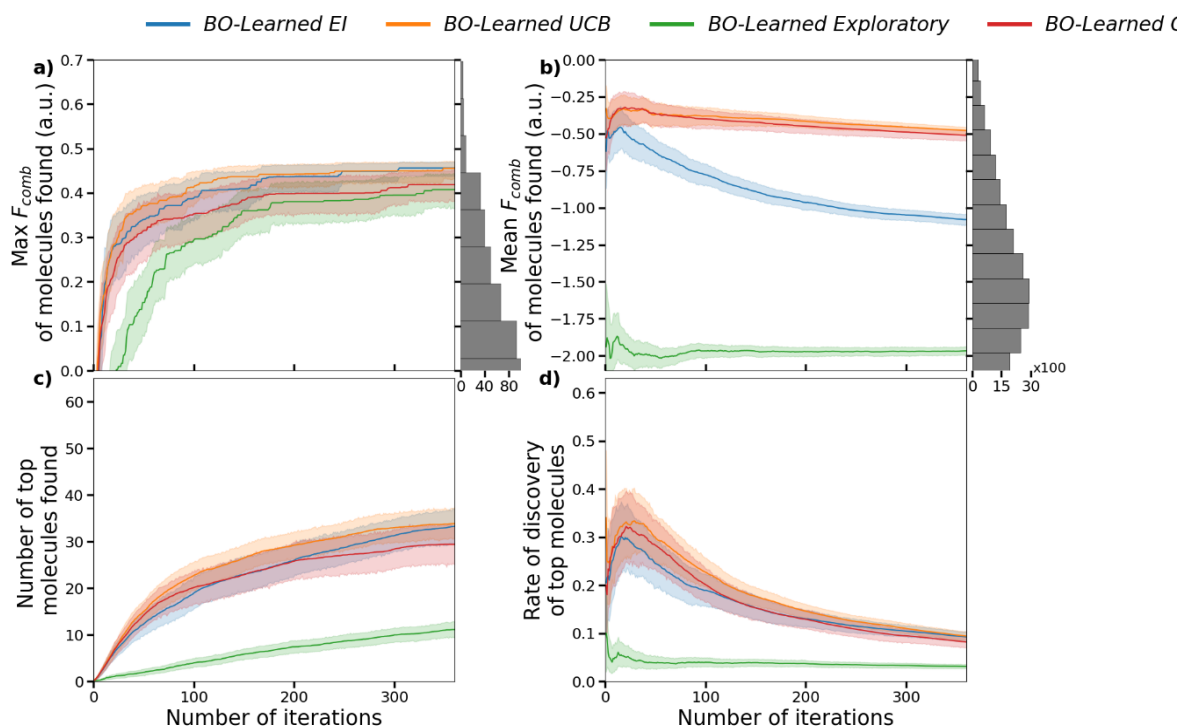

*Figure S26 Impact of the acquisition function on the search algorithm performance. Here we consider four different acquisition function: Expected improvement (EI), Upper confidence bound (UCB), Exploratory (element with max uncertainty), Greedy (element with maximum of the predictive mean). Performance of the search algorithms over different metrics when exploring the space of precalculated oligomers (~30000 unique oligomers). The solid-coloured lines show the mean $F_{comb}$ over 25 runs with different initial populations and the coloured shaded area shows the variance of the $F_{comb}$ over those different runs; a) Maximum $F_{comb}$ found in an oligomer up to the current iteration. The histogram on the right shows the distribution of the molecules in the benchmark dataset; b) Mean $F_{comb}$ of the oligomer found up to the current iteration; c) Number of oligomers in the top 1% found up to the current iteration (top 1% is 300 molecules); d) Discovery rate of the top 1% oligomers in the dataset, calculated as the (number of top oligomers found)/(number of iterations).*

Figure S25, shows the performance of the *BO-Learned* search algorithm over the benchmark search space with different acquisition functions. Here we consider four different acquisition function: Expected improvement (EI) which is used throughout the paper. Upper confidence bound (UCB) which is a sum of the predictive mean and the predicted uncertainty for a specific oligomer. Exploratory acquisition which only considers the predicted uncertainty and the Greedy approach which only considers the predictive mean. The Greedy and UCB acquisition perform similarly in this case, showing overall similar performance over the different metrics. The exploratory acquisition function suggests the most diverse set of oligomers but is not guided by the aim of the search to find an oligomer with high combined property. The Expected improvement in this case seems to balance exploration and exploitation in this case, which results in the highest number of top 1% oligomers found after 500 iterations.

# Results of the search algorithm over the unrestricted space.

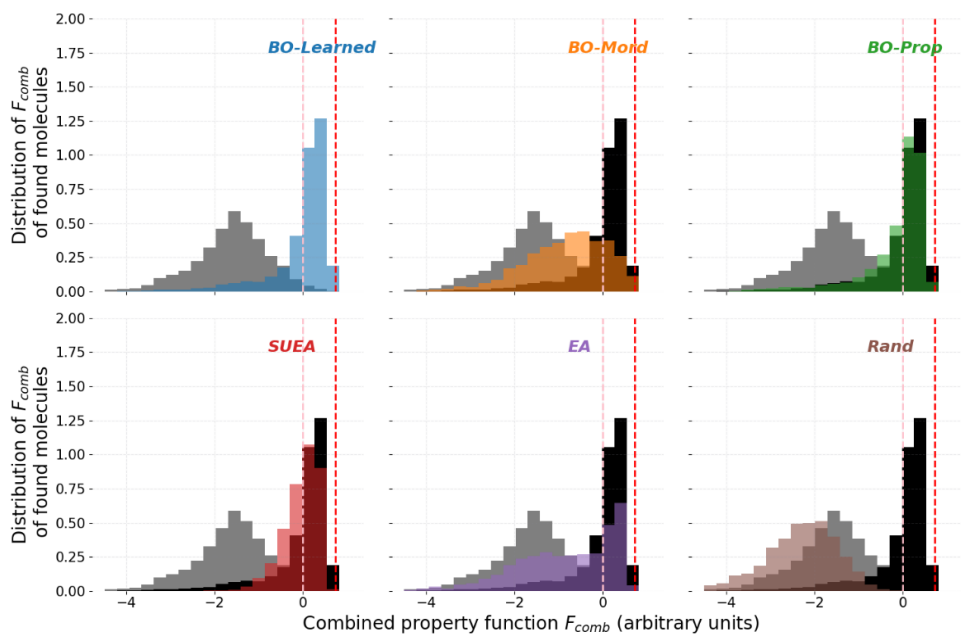

*Figure S27 Distribution of the target property of the oligomers suggested by each search algorithm for the first set of runs. the first set of runs are the 50 parallel searches done after the benchmark dataset. The grey shaded area shows the distribution of the data in the benchmark dataset. The black distribution in the panels shows the distribution from the BO-Learned search algorithm for comparison. The pink dashed line shows the threshold to have target property above 0 and the red dashed line shows the best element in the benchmark.*

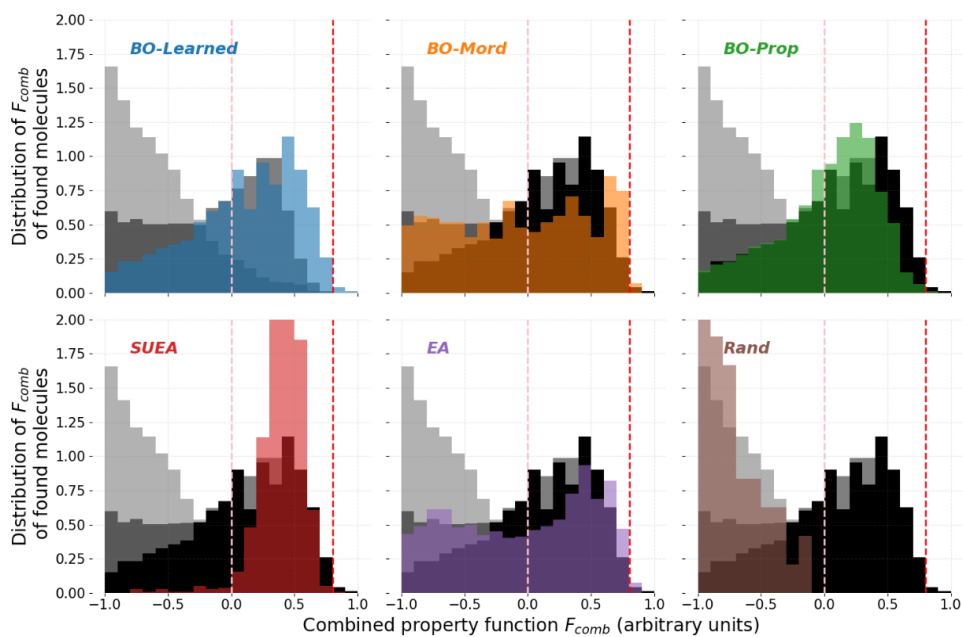

*Figure S27 Distribution of the target property of the oligomers suggested by each search algorithm for the second set of runs. the second set of runs, are the 20 parallel runs of the search approach which uses the dataset generated after the first run. The light grey shaded area shows the distribution of the data in the benchmark dataset, the grey shaded area shows the distribution of the data in the dataset from the first set of runs. The black distribution in the panels shows the distribution from the BO-Learned search algorithm for comparison. The pink dashed line shows the threshold to be among the top 1% of the oligomers in the starting dataset and the red dashed line shows the best element in the starting dataset.*

In the second set of runs with results presented in Figure S27 for the *SU-EA*, the ratio of molecules suggested with combined property increased considerably compared to the first set of runs. This can be related to the overall improved performance of the surrogate model to predict the combined property of the molecules. For the *BO-Learned*, although the surrogate model used to generate the representation performs better at predicting the combined property (as supported by the performance of the *SU-EA*), it suggested less molecules with $F_{comb}$ above 0 as compared to its performance in the first set of runs. This could be explained by the exploratory nature of the acquisition function considered. In the context of the expected improvement acquisition function, this means that the algorithm suggests elements with higher uncertainty rather than overall higher predicted combined property value. In terms of finding molecules that are either closer to the best of the current dataset or even better, the *BO-Mord* and the simple *EA*, show the best performance overall. In this case, the *EA* finds 7 molecules with higher combined property than the ones in the dataset, whereas the BO-*Learned* only finds 4 new molecules. These results confirm the initial observation, that a model that has more information about the space, fails to find elements better than the dataset it has been trained on.

# Computation time of searching over unrestricted space.

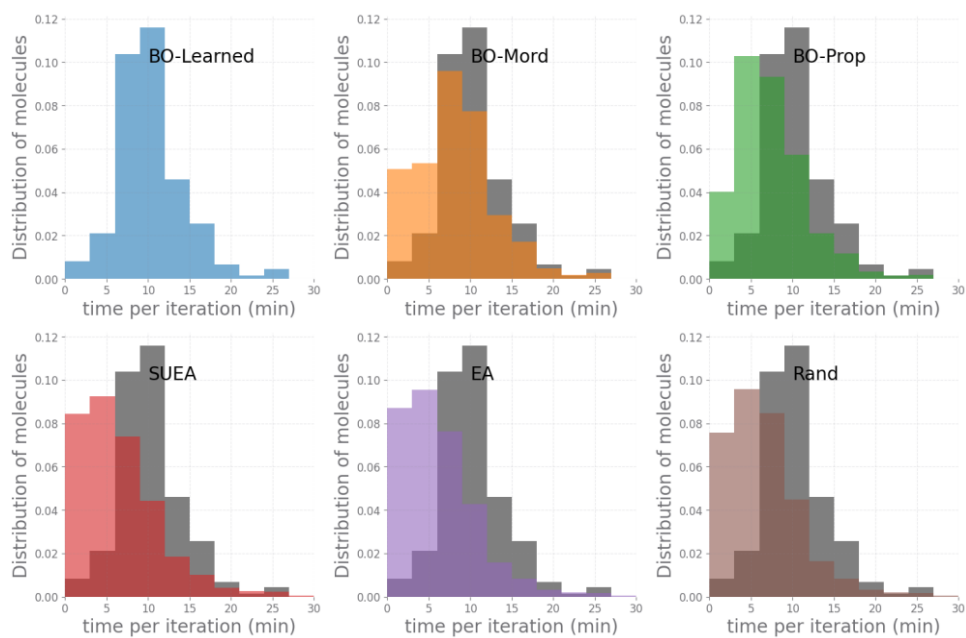

*Figure S29 Distribution of time per iteration for the different search algorithms. The distribution here is over the oligomers found through the second set of runs over the unrestricted space.*

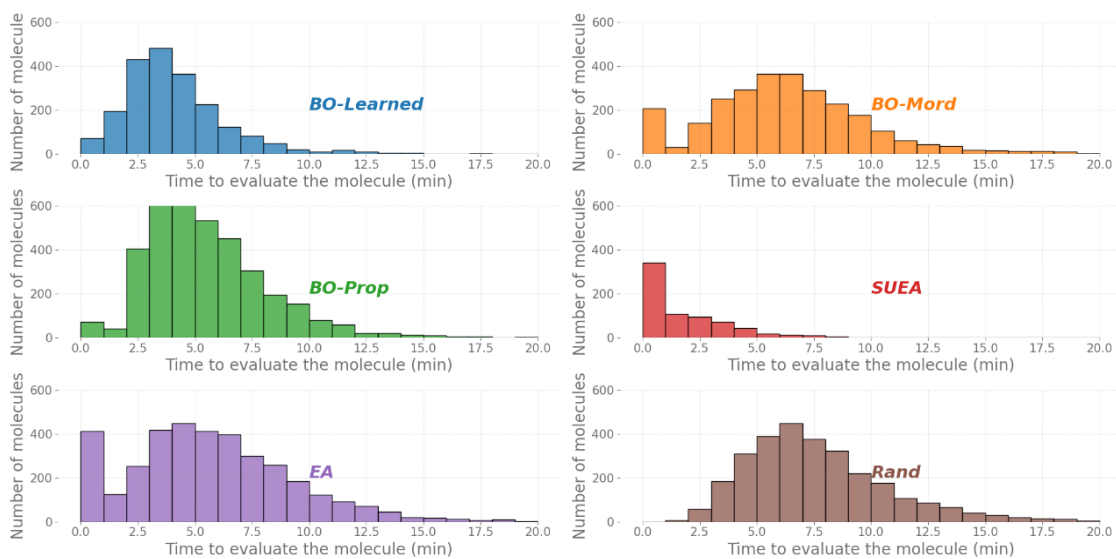

*Figure S30 distribution of the time needed to evaluate an oligomer using the evaluation function.*

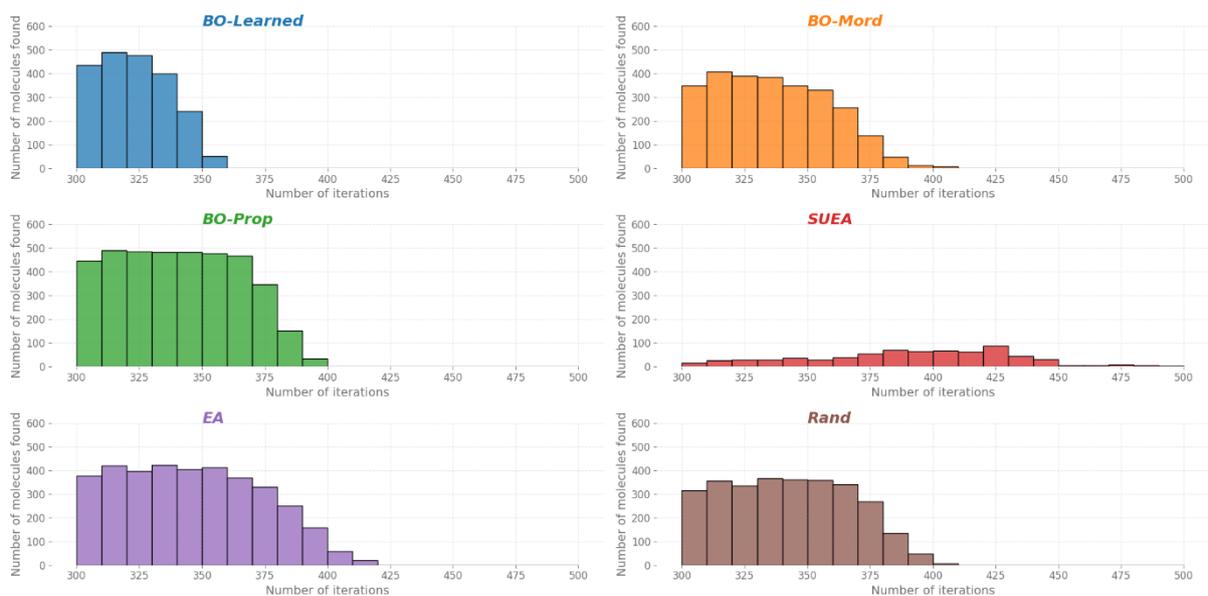

*Figure S31 New oligomers found over the set of runs across different number of iterations. Here we show the sum of the oligomers suggested by the parallel runs at a specific iteration. The change in the number of oligomers found with iteration is due to both a change in the number of runs that reached a certain number of iteration as well as the possibility of the parallel runs to suggest the same molecules.*

# DFT/TDDFT results

In this section, we show the results of calculating the properties of interest, namely the first excited state energy and oscillator strength as well the ionisation energy of the most promising molecules suggested by the different algorithms using a higher level of theory than XTB/sTDA. We consider the top 5 molecules suggested by the first run and the top 5 molecules suggested by the second run on the full search space. The structures of the molecules considered are shown in figure S32.

To calculate the properties of the molecules, we start with the geometries generated using the Supramolecular Toolkit (STK). These geometries are then optimized using the GFN2-xTB method, which is suitable for rapid and accurate electronic structure calculations of large molecular systems. Next, we use Gaussian to optimize the molecules in vacuum employing the $\omega$B97XD functional with the 6-31G* basis set. The $\omega$B97XD functional is chosen for its ability to accurately describe weak interactions and its proven correlation with experimental data, while the 6-31G* basis set provides a good balance between computational cost and accuracy.[78]

For ionization potential (IP) calculations, we optimize the geometry of the molecules in their cationic state. The IP is estimated by calculating the difference in total energy between the ground state geometry and the optimized cation geometry. To determine the excited state energy and oscillator strength, we perform time-dependent density functional theory (TD-DFT) calculations on the molecule in its ground state optimized geometry.

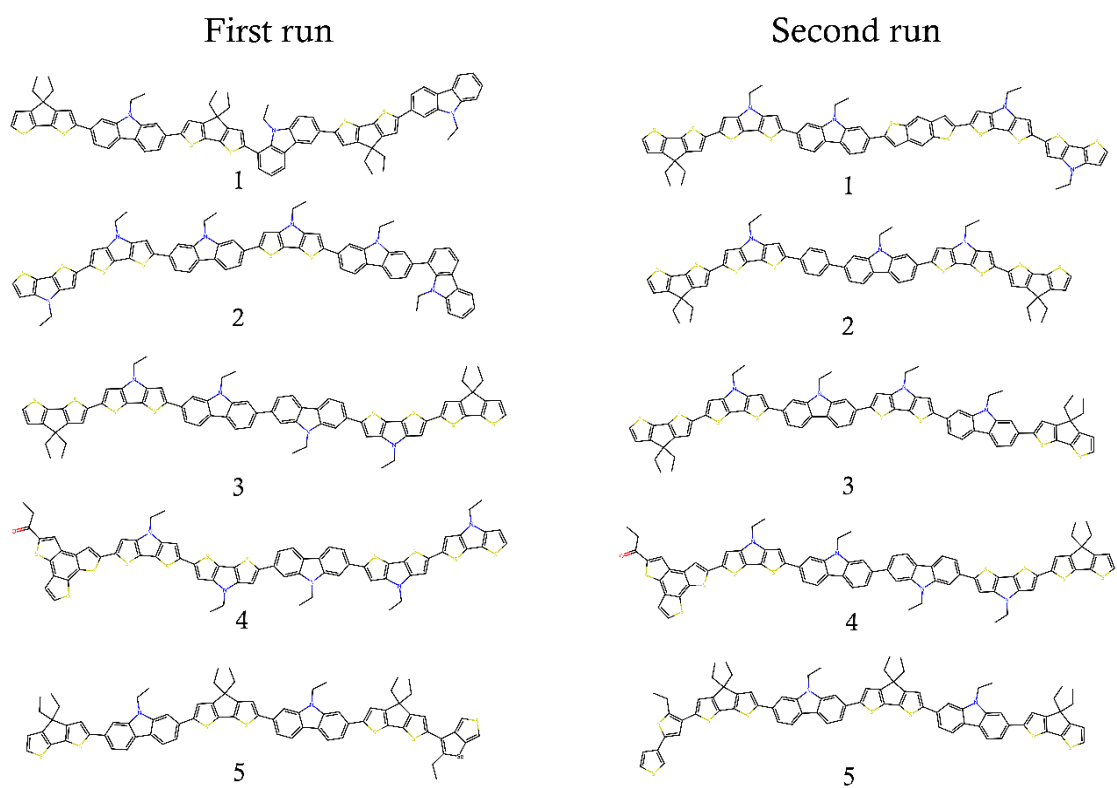

*Figure S 32 Best 5 molecules after each run considered for the DFT/TDDFT calculation. The labels shown here are the ones used in table S2.*

The comparison of the properties of interest for the ten selected molecules calculated at the XTB/sTDA level and the DFT/TDDFT level is shown in table S2. First most of the identified molecules using XTB/sTDA, are promising candidates for the application considered. Meaning the computed properties using DFT/TDDFT suggest that they would be good candidates as donor molecules with Y6. Specifically, they all show a considerably higher oscillator strength than other molecules considered for optoelectronic applications.[55] The differences observed between the calculated properties using the two methods is expected, however we observe that both $E_{S_1}$ and IP are within 0.3-0.5 eV of each other, confirming that XTB/sTDA is indeed a good proxy to compute the properties considered. However, a more thorough analysis of the correlation between the two levels of theory, along with experimental validation, is necessary to further confirm the potential of the proposed method for discovering new molecules suitable for real-world applications.

*Table S 2 Comparison of the XTB/sTDA and the DFT/TDDFT calculated properties of the 5 top performing molecules in each run.*

|  | XTB/sTDA | | | DFT/TDDFT | | |
| --- | --- | --- | --- | --- | --- | --- |
| Id | $E_{S_1}$ (eV) | IP (eV) | $f_{osc,1}$ | $E_{S_1}$ (eV) | IP (eV) | $f_{osc,1}$ |
| Second Run | | | | | | |
| 1 | 3.02 | 5.48 | 8.82 | 2.81 | 5.53 | 3.19 |
| 2 | 2.99 | 5.50 | 7.91 | 2.87 | 5.82 | 3.43 |
| 3 | 3.00 | 5.46 | 7.77 | 2.94 | 5.86 | 4.17 |
| 4 | 2.97 | 5.51 | 7.61 | 2.79 | 5.49 | 3.59 |
| 5 | 2.99 | 5.49 | 7.09 | 3.39 | 6.44 | 5.41 |
| First Run | | | | | | |
| 1 | 3.02 | 5.51 | 6.87 | 3.37 | 6.12 | 4.69 |
| 2 | 3.06 | 5.50 | 7.32 | 3.39 | 5.97 | 4.79 |
| 3 | 3.02 | 5.45 | 7.35 | 2.82 | 5.78 | 4.48 |
| 4 | 2.96 | 5.44 | 7.53 | 3.30 | 5.80 | 5.51 |
| 5 | 2.98 | 5.52 | 6.55 | 3.53 | 5.89 | 4.96 |